\documentclass[lefteqn,showpacs,preprint]{revtex4}

\usepackage{graphicx}
\usepackage{dcolumn}
\usepackage{bm}

\begin{document}
\draft
\title{Scattering mechanism in a step-modulated subwavelength metal slit: a multi-mode multi-reflection analysis}
\author{Chao Li$^{1,2}$, Yun-Song Zhou$^{2,\dag}$, Huai-Yu Wang$^{3,\ddag}$, and Jian-Hong Guo$^{2}$}
\date{\today }
\draft

\address{$^1$ School of Mathematical Sciences, Capital Normal University, Beijing 100048 China}

\address{$^2$ Center of Theoretical Physics, Department of Physics, Capital Normal University. Beijing
100048, China}

\address{$^3$ Department of Physics, Tsinghua University, Beijing 100084, China}

\begin{abstract}
In this paper, the scattering/transmission inside a step-modulated
subwavelength metal slit is investigated in detail. We firstly
investigate the scattering in a junction structure by two types of
structural changes. The variation of transmission and reflection
coefficients depending on structural parameters are analyzed. Then a
multi-mode multi-reflection model based on ray theory is proposed to
illustrate the transmission in the step-modulated slit explicitly.
The key parts of this model are the multi-mode excitation and the
superposition procedure of the scatterings from all possible modes,
which represent the interference and energy transfer happened at
interfaces. The method we use is an improved modal expansion method
(MEM), which is a more practical and efficient version compared with
the previous one [Opt. Express \textbf{19}, 10073 (2011)]. In
addition, some commonly used methods, FDTD, scattering matrix
method, and improved characteristic impedance method, are compared
with MEM to highlight the preciseness of these methods.
\end{abstract}


\maketitle

\newpage

\section{Introduction}

Subwavelength metal slits, as a kind of metal/insulator/metal
waveguides, have attracted much attention in recent years not only
because of their ability to guide light beyond the diffraction
limit, but also because of several remarkable advantages, such as
strong field localization, simplicity, and convenience for
fabrication and integration into optical circuits [1-14]. When light
(infrared, visible spectrum) propagates along a metal/air interface,
it will excite a collective oscillation of free electrons at the
surface of the metal, causing a field exponentially decaying away
from the interface. This mode is called as surface plasmon polariton
(SPP) [2-6]. In a subwavelength metal/air/metal slit, the case is
somehow different, since the in-slit SPP [5] wave decays
exponentially in the metals and is flat in the air, which is the
lowest eigenmode in the slit structure and the core part of the
subwavelength metallic optics.

Step modulation is one of the key elements in photonic engineering
that are employed in subwavelength metal structures to design and
fabricate functional plasmonic devices, such as filters [7-10],
reflectors [11], and photonic bandgap structures [11,12]. Besides,
the step modulation is of important theoretical significance since
they are helpful for investigating SPP scattering. Had the knowledge
and combined with the staircase approximation and transfer matrix
technique, numerical results of more complicated structures can be
obtained.

Up to now, a number of methods have been used to calculate the SPP
scattering/transmission inside a step-modulated slit. The
finite-difference time-domain method (FDTD) is a well developed
simulation method that provides relatively accurate results and has
been considered as a standard for testing other theoretical methods
[5-14]. The effective index method, on the other hand, is a
simplified and direct theoretical method where only the SPP modes
are involved in the calculation, a method of one mode approximation;
however, this simplification causes the loss of the scattering
information considerably and the numerical imprecision turned out to
be considerable under some conditions [6]. Matsuzaki \emph{et al}.
presented a transmission model and gave a better description of the
SPP scattering using the characteristic impedance method [13].
Pannipitiya \emph{et al}. [14] suggested an improved version of this
method, which will be called as improved characteristic impedance
method (ICIM) in this paper. Lin \emph{et al}. presented a similar
transmission model and used the scattering matrix method (SMM) to
calculate the transmission [7-9]. Although the calculated results
from these two methods fit the FDTD results, two approximations, one
mode approximation and quasi-statistic approximation, are used in
the calculation, which limits their application scope and numerical
precision, as will be discussed in Sec. 3 below. Recently, we
successfully applied the modal expansion method (MEM) in discussing
the wave behavior inside a symmetric step-modulated slit [6]. This
method did not involve the two mentioned approximations and provided
more accurate results.

In this paper, the MEM is further improved so as to apply to the
asymmetric modulated case in investigating the
scattering/transmission mechanisms inside a step-modulated slit. A
multi-mode multi-reflection model is proposed to explain the
transmission process. A remarkable advantage of MEM is that its
precision is controllable. This enables us to discuss the
preciseness of FDTD and ICIM by comparing the results from these
methods and MEM.

The paper is arranged as follows. Section 2 sets our model of a
step-modulated metal slit and presents the improved MEM formulas. In
Sec. 3, the scattering in a junction structure is studied firstly as
a prerequisite for later discussion, and then a multi-mode
multi-reflection model is proposed to reveal the transmission
mechanism in a step-modulated slit. Comparisons between MEM, FDTD,
SMM, and ICIM are also given in this section to highlight the
restriction of the one mode approximation and quasi-statistic
approximation. Finally, conclusions are presented in Sec. 4.

\section{Model and the improved MEM}

In this section, we set the model of single-slit structure and
present the formulas of the improved MEM which is more practical and
efficient compared with that in our previous work [6].

\begin{figure}[htbp]
\centering\includegraphics  [width=8cm] {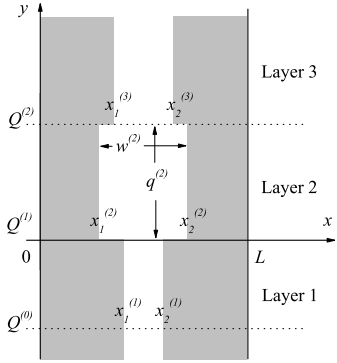} \caption{Sketch of a
step-modulated metal slit structure confined in $x$ direction with
perfectly conducting walls at 0 and $L=2$ $\mu$m. A TM wave with
wavelength $\lambda_{0}$ is normally launched at $y=Q^{(0)}$.
$Q^{(1)}=0$.}
\end{figure}

The model structure is shown in Fig. 1. It is infinitely large in
$yz$ plane but confined in $x$ direction by perfectly conducting
walls at $x=0$ and $x=L$. The structure is divided into three
regions along $x$ direction, composed of silver/air/silver, and
three layers along $y$ direction. The lower boundary of the $l$th
layer is labeled as $Q^{(l-1)}$, and the interfaces between regions
are by $x_{1}^{(l)}$ and $x_{2}^{(l)}$. The denotations $q^{(l)}$
and $w^{(l)}$ label layer height and slit width, respectively. A
transverse magnetic (TM) wave, with magnetic field $\textbf{H}$
being in $z$ direction, is normally launched from $y=Q^{(0)}$ in
Layer 1 and propagates upward. The structural parameters are given
in the caption of Fig. 1.

The dielectric constant of silver as a function of the wavelength of
the incident wave $\lambda_{0}$ is evaluated as
$\varepsilon_{Ag}=(3.57-54.33\lambda_{0}^{2})+i(-0.083\lambda_{0}+0.921\lambda_{0}^{3}
)$ by fitting the experimental data [15], which is valid for
$0.6\leq \lambda_{0}\leq1.6$ $\mu$m. In this paper, the wavelength
is mainly set as $\lambda_{0}=1$ $\mu$m; thus,
$\varepsilon_{Ag}=-50.76+0.083i$.

In the remaining part of this section, we suggest an improved
version of the MEM, which has the same output as the previous one
[6] but is easier and faster.

The substance of MEM is to expand the unknown functions
(electromagnetic field distribution in present case) by a complete
set of orthogonal functions. This makes the MEM has two folds: one
is the eigenvalue problem of the system; the other is to establish
and solve the coupled equations subject to boundary conditions.
However, the choice of the complete set in modal expansion is not
unique, but depends on the configuration of the system. It can be
the eigenfunctions of a specific structure or other functions such
as sine or exponential functions.

In Ref. [6], the fields in the given structure were handled by
separation of variables. The factors containing $x$ variable were
expanded by the eigenmodes $\{\psi_{n}^{(l)}(x)\}$ between two
perfectly conducting walls. The magnetic fields were expressed as
\begin{equation}
H_{z}(x,y)=\left\{
\begin{array}{lcl}
\sum_{n}\psi_{n}^{(1)}(x)\left[I_{n}e^{ik_{yn}^{(1)}\left(y-Q^{(0)}\right)}+R_{n}e^{-ik_{yn}^{(1)}\left(y-Q^{(1)}\right)}\right], & & {Q^{(0)}\leq y<Q^{(1)}}\\
\sum_{n}\psi_{n}^{(2)}(x)\left[E_{n}e^{ik_{yn}^{(2)}\left(y-Q^{(1)}\right)}+F_{n}e^{-ik_{yn}^{(2)}\left(y-Q^{(2)}\right)}\right], & & {Q^{(1)}\leq y<Q^{(2)}}\\
\sum_{n}\psi_{n}^{(3)}(x)T_{n}e^{ik_{yn}^{(3)}\left(y-Q^{(2)}\right)},
& & {Q^{(2)}\leq y<\infty,}
\end{array} \right.
\end{equation}
where $I_{n}$, $R_{n}$, $E_{n}$, $F_{n}$, and $T_{n}$ were expansion
coefficients which involved the scattering/transmission information
of every eigenmode. It was inevitable to solve a transcendental
equation in order to achieve the eigenvalues and eigenfunctions.
Even with assistance of a powerful root-seeking method [16], this
procedure was still time-consuming. Moreover, each layer had its own
eigenfunctions. At an interface, it was required by the boundary
conditions to calculate the overlap between the eigenfunctions at
the two sides of the interface, called as coupling integrals. Such
integrals brought complexity to the program.

To avoid these difficulties, in this paper the factors containing
$x$ variable were expanded by a sine basis subject to the perfectly
conducting boundary condition. That is to say, the complete set
$\{\varphi_{n}(x)\}$ is chosen as
\begin{equation}
\varphi_{n}(x)=\sqrt{2/L}\sin(k_{xn}x), \ \ k_{xn}=n\pi/L, \ \ n=1,
2, 3\cdot\cdot\cdot.
\end{equation}
The eigenvalues $k_{xn}$ are solely determined by the distance
between the two perfectly conducting walls, independent of the
positions $x_{1}^{(l)}$ and $x_{2}^{(l)}$, so that is valid for all
the three layers.

Correspondingly, the magnetic fields and its derivative in the three
layers can be expressed as [17]
\begin{equation}
H_{z}(x,y)=\left\{
\begin{array}{lcl}
\sum_{n}\varphi_{n}(x)\sum_{m}W_{n,m}^{(1)}\left[i_{m}e^{ik_{ym}^{(1)}\left(y-Q^{(0)}\right)}+r_{m}e^{-ik_{ym}^{(1)}\left(y-Q^{(1)}\right)}\right], & & {Q^{(0)}\leq y<Q^{(1)}}\\
\sum_{n}\varphi_{n}(x)\sum_{m}W_{n,m}^{(2)}\left[e_{m}e^{ik_{ym}^{(2)}\left(y-Q^{(1)}\right)}+f_{m}e^{-ik_{ym}^{(2)}\left(y-Q^{(2)}\right)}\right], & & {Q^{(1)}\leq y<Q^{(2)}}\\
\sum_{n}\varphi_{n}(x)\sum_{m}W_{n,m}^{(3)}t_{m}e^{ik_{ym}^{(3)}\left(y-Q^{(2)}\right)},
& & {Q^{(2)}\leq y<\infty,}
\end{array} \right.
\end{equation}
where $i_{m}$, $r_{m}$, $e_{m}$, $f_{m}$, and $t_{m}$ are the
expansion coefficients. The insertions of Eq. (3) into Helmholtz
equation yields an eigenvalue problem in each layer expressed by
$A^{(l)}W^{(l)}=(ik_{y}^{(l)})^{2}W^{(l)}$ with $ik_{y}^{(l)}$ and
$W^{(l)}$ being the eigenvalues and eigenfunctions, and the operator
$A^{(l)}$ being [18]
\begin{equation}
A^{(l)}=-\left[\tilde{E}^{(l)}\right]^{-1}\left\{k_{0}^{2}[I]+[K]\left[E^{(l)}\right]^{-1}[K]\right\},
\end{equation}
where
\begin{equation}
\left\{
\begin{array}{lcl}
\left[\tilde{E}^{(l)}\right]_{mn}=\int_{0}^{L}\varphi_{m}(x)\varphi_{n}(x)/\varepsilon^{(l)}(x)dx\\
\left[E^{(l)}\right]_{mn}=\int_{0}^{L}\varphi_{m}(x)\varphi_{n}(x)\varepsilon^{(l)}(x)dx\\
\left[K\right]_{mn}=\int_{0}^{L}\varphi_{m}(x)\frac{\partial}{\partial x}\varphi_{n}(x)dx\\
\left[I\right]_{mn}=\delta_{mn}.
\end{array} \right.
\end{equation}
In these equations , $[\cdot]$ denotes a $N\times N$ matrix where
$N$ is the truncation number, and $k_{0}=2\pi/\lambda_{0}$ is the
wave vector in vacuum.

Here we mention the two advantages of the sine expansion in $x$
direction. One is that the eigenvalue problem, Eq. (4), is very easy
for computer implementation, which avoids the cumbersome
solution-seeking procedure necessary in the eigenmode expansion [6].
The other is that the complete set of sine functions are the same
for all layers, so that the coupling integrals at the interfaces
become quite simple. These two advantages make the calculation
program greatly simplified.

Corresponding to Eq. (3), the derivative of the magnetic field is
expressed as
\begin{equation}
\frac{1}{\varepsilon}\frac{\partial}{\partial y}H_{z}(x,y)=\left\{
\begin{array}{lcl}
\sum_{n}\frac{\varphi_{n}(x)}{\varepsilon^{(1)}}\sum_{m}W_{n,m}^{(1)}ik_{ym}^{(1)}\left[i_{m}e^{ik_{ym}^{(1)}\left(y-Q^{(0)}\right)}-r_{m}e^{-ik_{ym}^{(1)}\left(y-Q^{(1)}\right)}\right], & & {Q^{(0)}\leq y<Q^{(1)}}\\
\sum_{n}\frac{\varphi_{n}(x)}{\varepsilon^{(2)}}\sum_{m}W_{n,m}^{(2)}ik_{ym}^{(2)}\left[e_{m}e^{ik_{ym}^{(2)}\left(y-Q^{(1)}\right)}-f_{m}e^{-ik_{ym}^{(2)}\left(y-Q^{(2)}\right)}\right], & & {Q^{(1)}\leq y<Q^{(2)}}\\
\sum_{n}\frac{\varphi_{n}(x)}{\varepsilon^{(3)}}\sum_{m}W_{n,m}^{(3)}ik_{ym}^{(3)}t_{m}e^{ik_{ym}^{(3)}\left(y-Q^{(2)}\right)},
& & {Q^{(2)}\leq y<\infty,}
\end{array} \right.
\end{equation}
Applying the layer boundary conditions, we obtain the coupled
equations as follows:
\begin{equation}
\left\{
\begin{array}{lcl}
\sum_{m}W_{pm}^{(1)}\left[i_{m}e^{ik_{ym}^{(1)}q^{(1)}}+r_{m}\right]=\sum_{m}W_{pm}^{(2)}\left[e_{m}+f_{m}e^{ik_{ym}^{(2)}q^{(2)}}\right]\\
\sum_{n}\tilde{E}_{pn}^{(1)}\sum_{m}W_{nm}^{(1)}k_{ym}^{(1)}\left[i_{m}e^{ik_{ym}^{(1)}q^{(1)}}-r_{m}\right]=\sum_{n}\tilde{E}_{pn}^{(2)}\sum_{m}W_{nm}^{(2)}k_{ym}^{(2)}\left[e_{m}-f_{m}e^{ik_{ym}^{(2)}q^{(2)}}\right]\\
\sum_{m}W_{pm}^{(2)}\left[e_{m}e^{ik_{ym}^{(2)}q^{(2)}}+f_{m}\right]=\sum_{m}W_{pm}^{(3)}t_{m}\\
\sum_{n}\tilde{E}_{pn}^{(2)}\sum_{m}W_{nm}^{(2)}k_{ym}^{(2)}\left[e_{m}e^{ik_{ym}^{(2)}q^{(2)}}-f_{m}\right]=\sum_{n}\tilde{E}_{pn}^{(3)}\sum_{m}W_{nm}^{(3)}k_{ym}^{(3)}t_{m},
\end{array} \right.
\end{equation}

The incident coefficients $i_{m}$ are determined by the incident
wave. In this paper, the incident wave is always a SPP wave launched
in Layer 1, namely, $\psi_{1}^{(1)}(x)$. Therefore, one naturally
has
\begin{equation}
\int_{0}^{L}\varphi_{p}(x)\psi_{1}^{(1)}(x)dx=\sum_{m}W_{pm}^{(1)}i_{m},
\end{equation}
which determines the coefficients $i_{m}$. After setting the
incident coefficients $i_{m}$, the four groups of coefficients,
$r_{m}$, $e_{m}$, $f_{m}$, and $t_{m}$, can be obtain from Eq. (7).
Thus all the field quantities are obtained.

In this paper, we will focus on discussing the
reflection/transmission mechanisms, which are mainly presented by
the reflection coefficients $R_{n}$ and transmission coefficients
$T_{n}$ in Eq. (1). For example, the amplitudes of the SPP modes in
Layer 1 and 3, $|R_{n}|$ and $|T_{n}|$, are the reflection and
transmission efficiencies of the system, and their arguments,
$arg(R_{n})$ and $arg(T_{n})$, are the corresponding phase shifts.
Therefore, a projection between the fields calculated by Eq. (3) and
the eigenmodes $\{\psi_{n}^{(l)}(x)\}$ is implemented for obtaining
the $R_{n}$ and $T_{n}$. In the following, the absolute values of
these coefficients may generally be named as excitation efficiency.

Thus we accomplish our formulation presentation. This method is
briefly outlined as follows: the field is expanded by sine functions
which are complete and uniform in all layers. The corresponding
eigenvalue problem becomes a matrix form as shown in Eq. (4), which
makes the calculation quite easy. Accordingly, the procedure here is
much more practical and efficient compared to the previous one [6].
At last, the overlaps between the calculated field and the
eigenmodes in Eq. (1) give the required reflection and transmission
coefficients necessary for physical analysis.

Although we merely study the three-layer structure, our procedure
developed here is easily applied to more complicated structures by
implanting the S matrix algorithm [19] or the enhanced transmittance
matrix approach [20].

\section{Numerical results and analysis}

In this section, we investigate the scattering/transmission
mechanisms inside a step-modulated subwavelength metal slit. To do
so, the scattering in a junction structure is first discussed in
detail, since the slit comprises more than one junction structure.
Then we disclose the multi-mode multi-reflection model in the
transmission process in the slit. By the way, the numerical
precision of FDTD and ICIM is discussed by comparing results of
these two methods and MEM.

In calculation, the confinement is set as $L=2$ $\mu$m. We have
tested that $800$ modes, $N=800$ in Eq. (4), are enough to give
results with precision up to four significant digits. The
convergence test for truncation number $N$ and the preciseness test
for confined width $L$ will be carried out later in Fig. 7.

\subsection {Junction structures}

A junction structure is the connection of two half-infinitely long
slits with widths being denoted by $w^{(1)}$ and $w^{(3)}$,
respectively, which can be easily realized in Fig. 1 by setting the
height of Layer 2 to $0$. In Ref. [6], some scattering properties of
symmetric structures have been revealed. For example, the main
ingredient of the fields inside the slits were guided modes which
played a very important role in scattering/transmission, and the
unimportant components were the radiation modes excited which were
necessary to fulfill boundary condition, but had little contribution
to the transmission. So the following discussion will focus on the
guided modes. However, the discussion in Ref. [6] was limited to the
symmetric case. Here we present a detailed investigation on how the
scattering is affected by asymmetry.

Two types of structural changes are considered. In Type I, the
widths of the two slits are fixed and the position of the narrower
one can be anywhere between left to right, as shown in the inner
panel of Figs. 2(a) and (c). In Type II, the left walls of the slits
are aligned and the width of the narrower one is fixed, but that of
the wider one can vary, as schematically shown in the inner panel of
Figs. 3(a) and (c).

\begin{figure}[htbp]
\centering\includegraphics  [width=6cm] {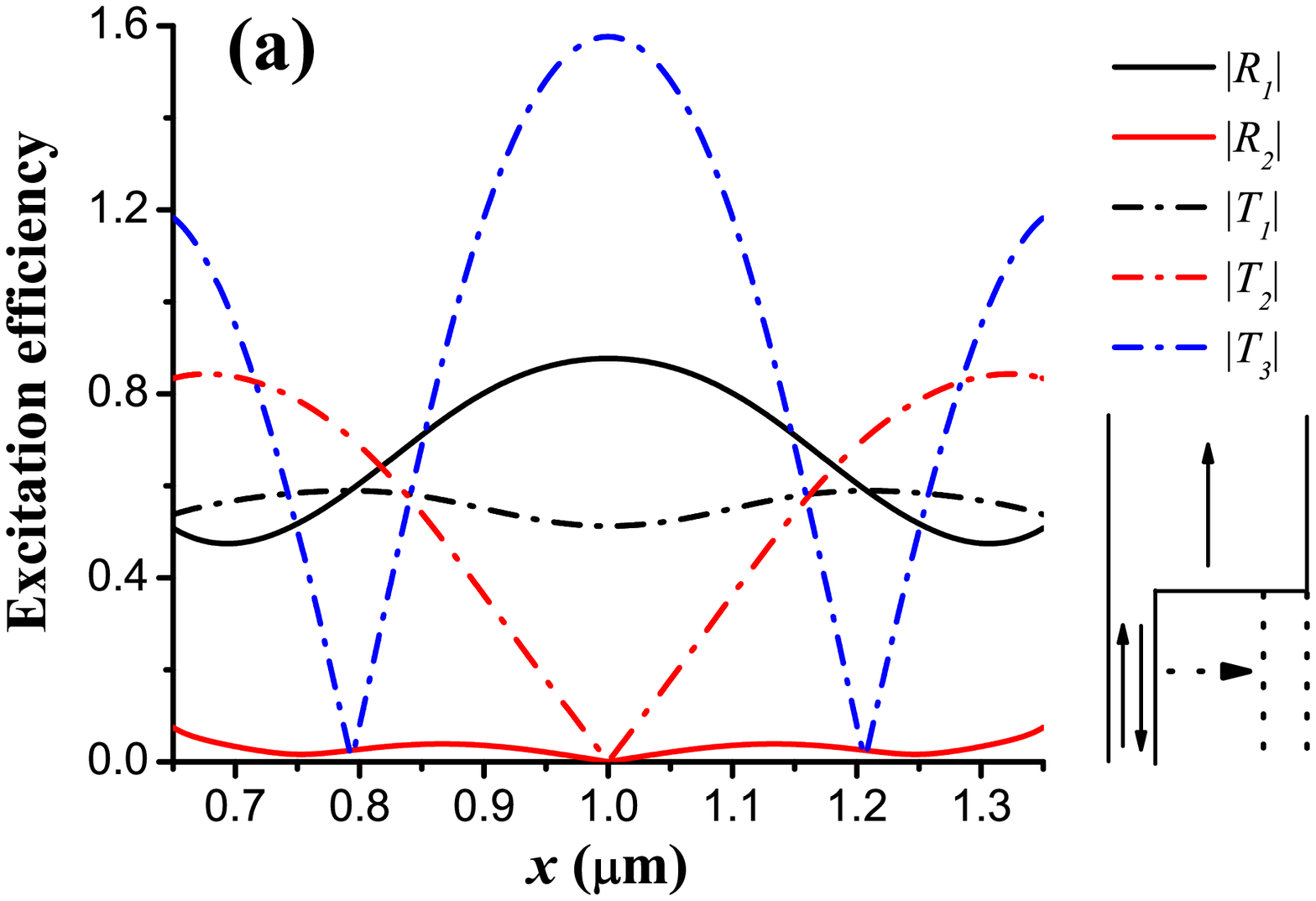}
\includegraphics  [width=6cm] {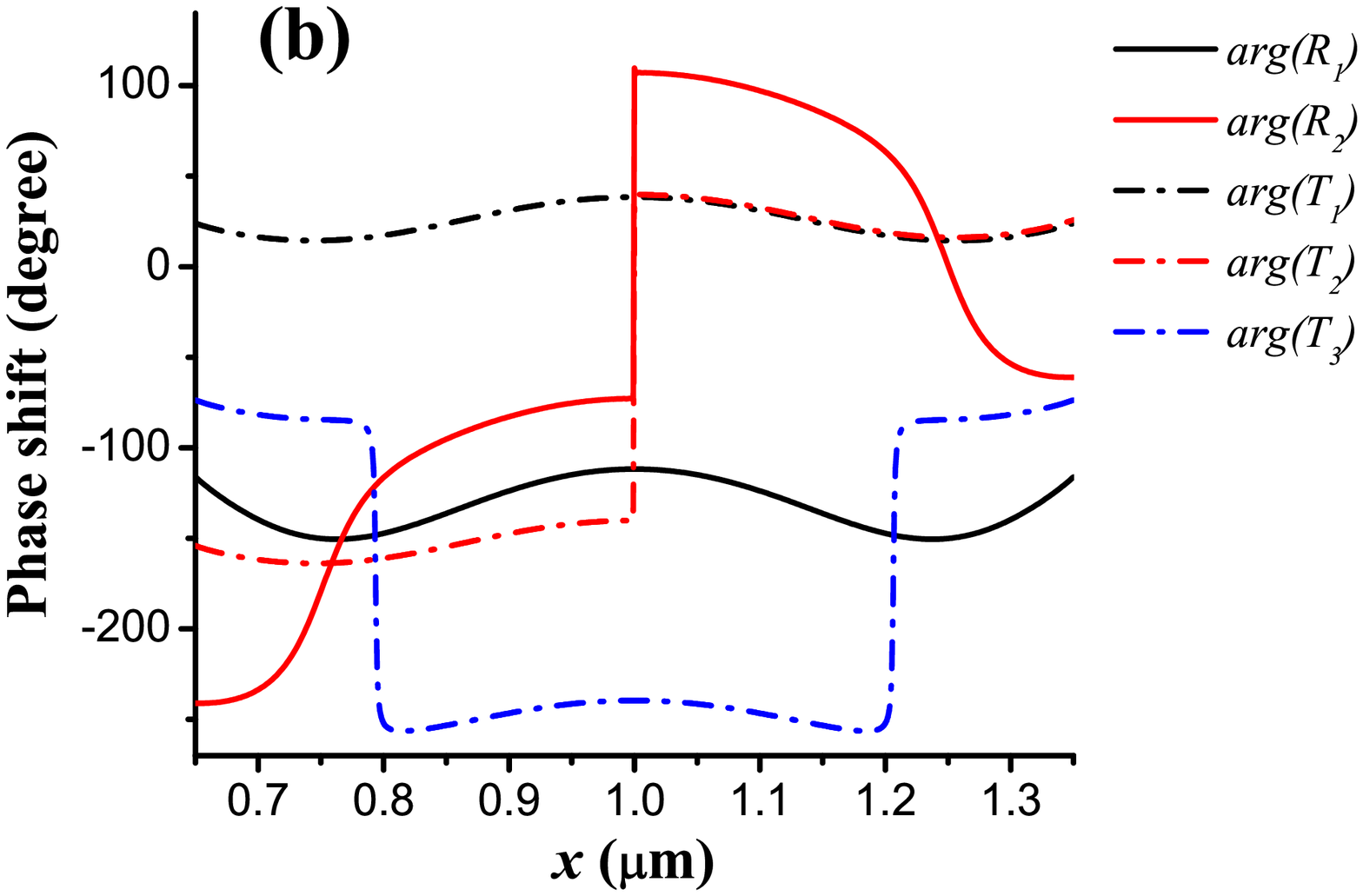}
\includegraphics  [width=6cm] {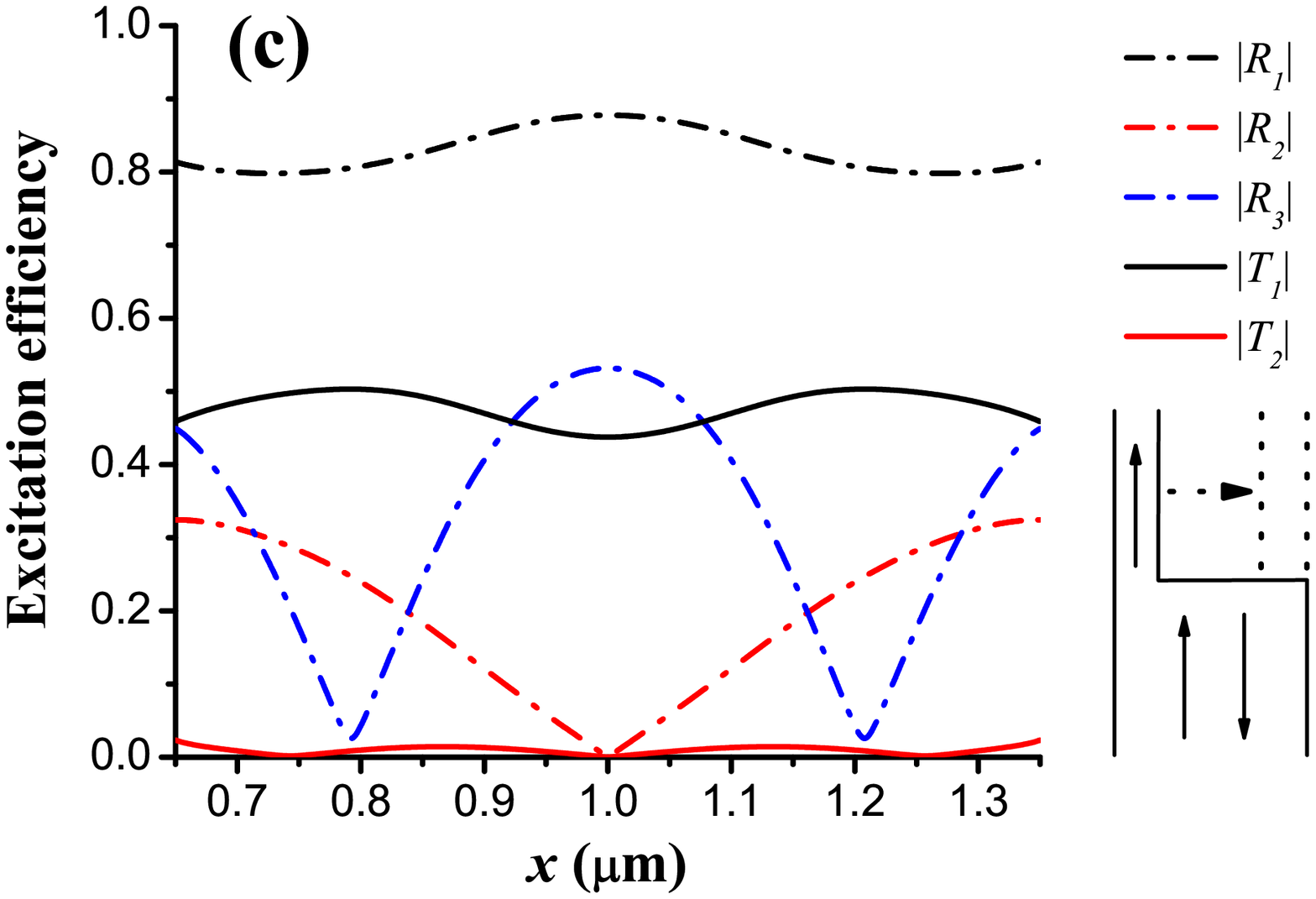}
\includegraphics  [width=6cm] {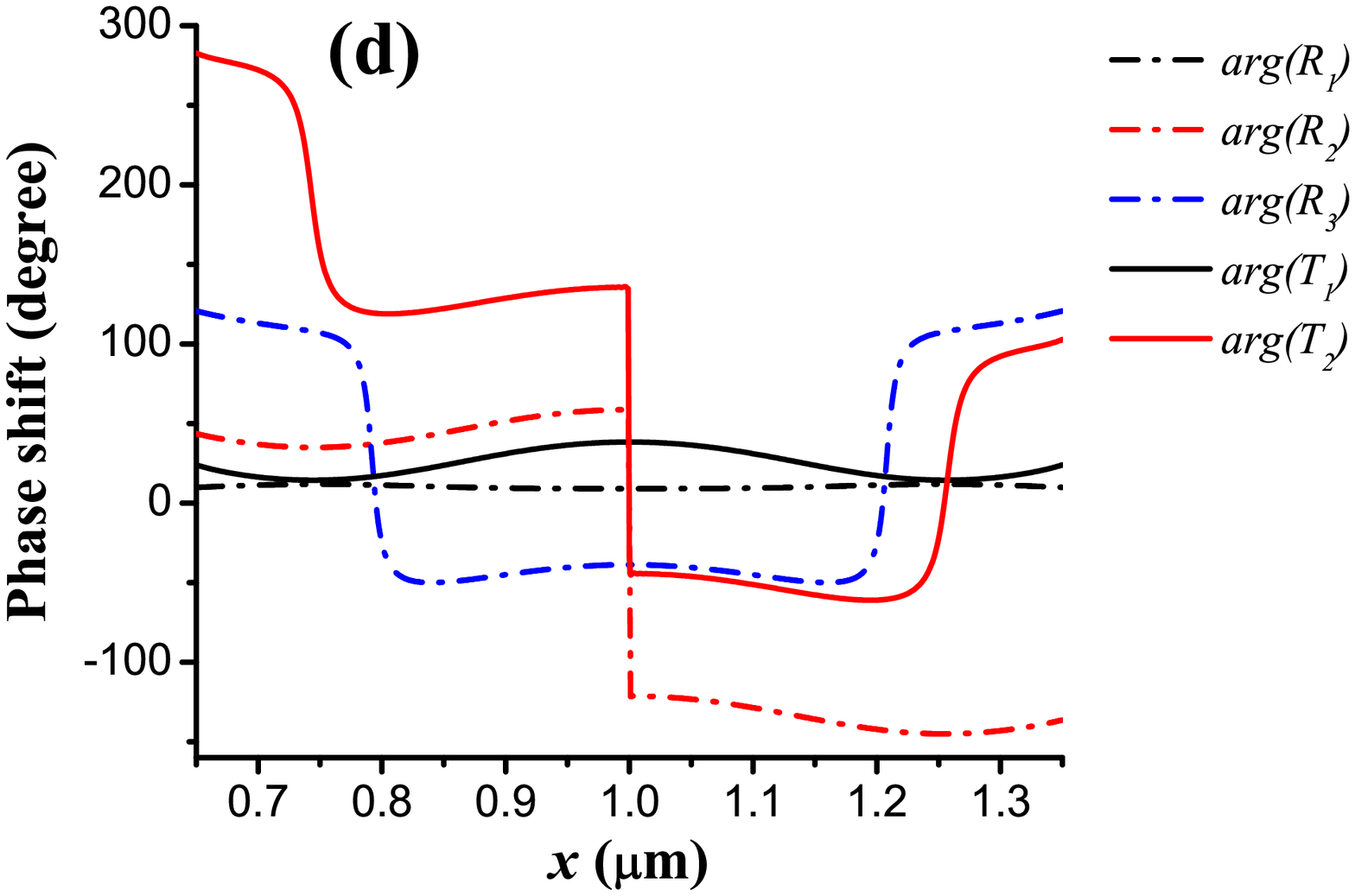}
\includegraphics  [width=6cm] {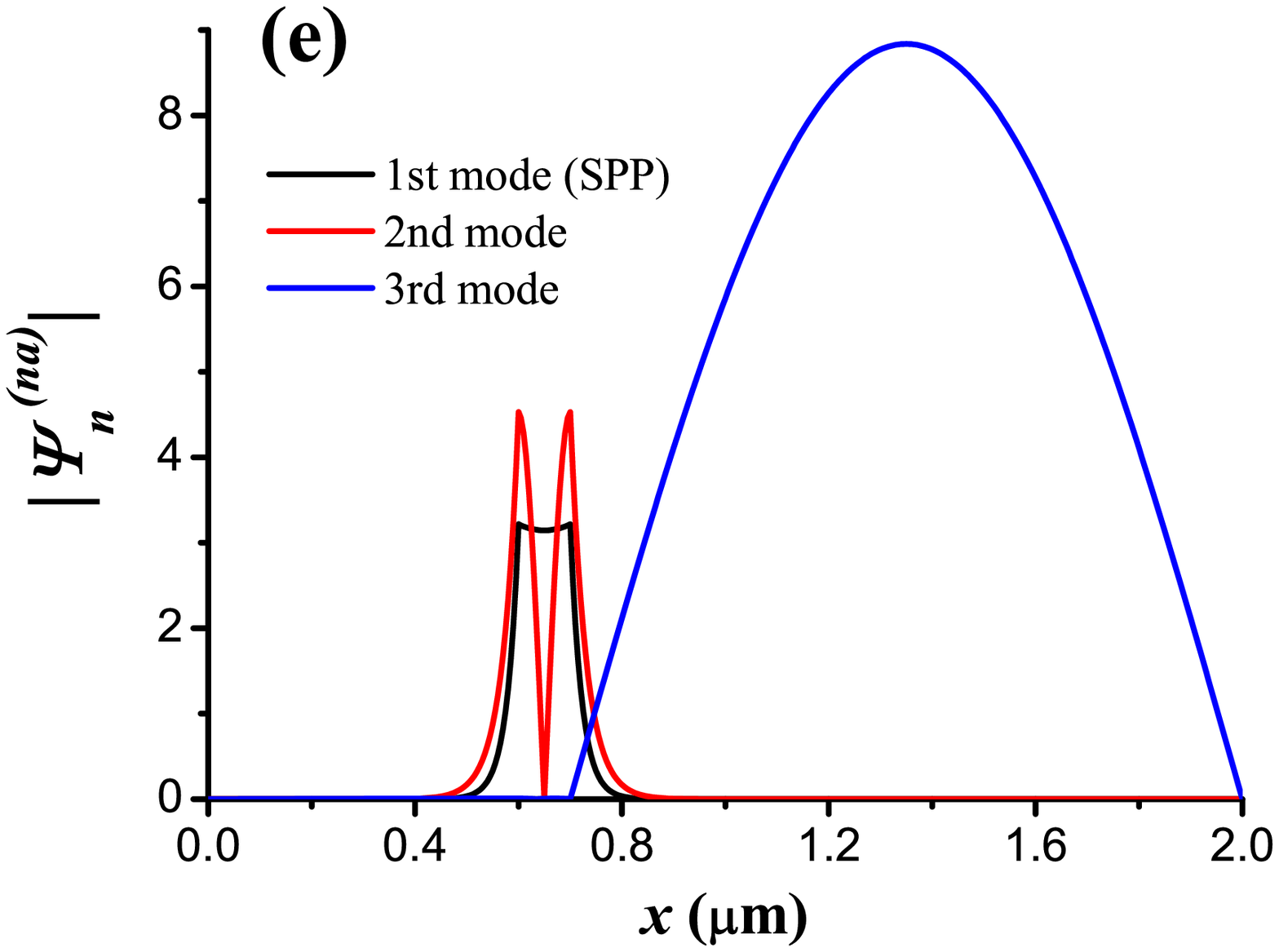}
\includegraphics  [width=6cm] {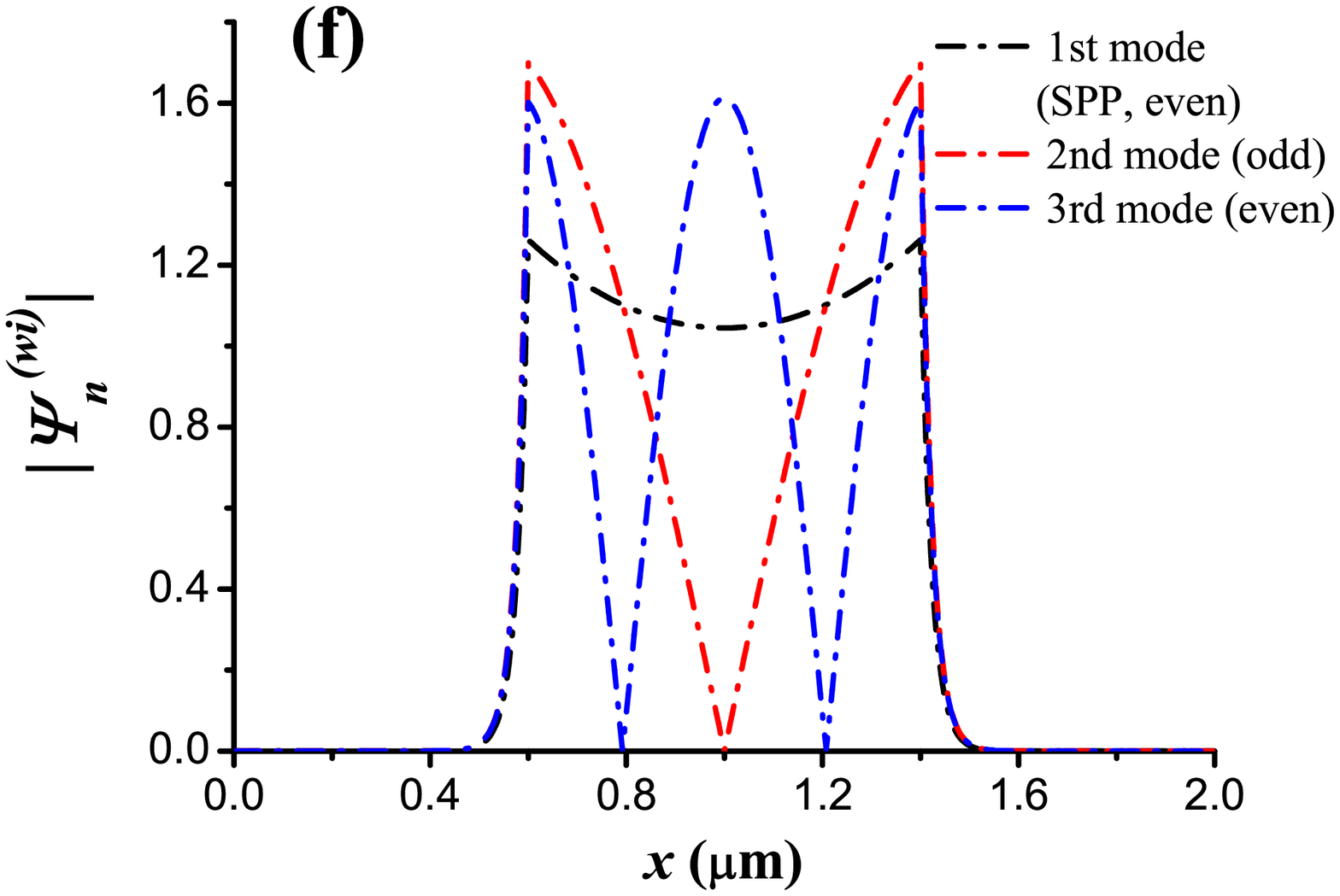}
 \caption{(color online). Scattering in Type I junction structures. The left
wall of the wider slit is fixed at $x_{1}=0.6$ $\mu$m.
$Q^{(0)}=Q^{(1)}=Q^{(2)}=0$. In all the figures, the results of the
wider slit are plotted by solid lines and those of the narrower slit
by dash-dotted lines. The narrower slit moves from the left to
right. In Figs. (a) to (d), the $x$-axes are its central position.
For the structure with $w^{(1)}=0.1$ and $w^{(3)}=0.8$ $\mu$m, (a)
excitation efficiency; (b) phase shift. For the structure with
$w^{(1)}=0.8$ and $w^{(3)}=0.1$ $\mu$m, (c) excitation efficiency;
(d) phase shift. In (e) and (f) plotted are the absolute value of
the eigen functions in the narrower and wider slits, respectively,
when their left walls are aligned.}
\end{figure}

The results of Type I structure are plotted in Fig. 2. The left wall
of the wider slit is at $x_{1}=0.6$ $\mu$m. For  $w^{(1)}=0.1$ and
$w^{(3)}=0.8$ $\mu$m, when the position of narrower slit is moved
from the left to right, the excitation efficiencies and their phase
shifts are plotted in Figs. 2(a) and (b), respectively. In Figs.
2(c) and (d) are the excitation efficiencies and their phase shifts
of structure $w^{(1)}=0.8$ and $w^{(3)}=0.1$ $\mu$m. In Figs. 2(e)
and (f) are the absolute values of the first three eigenfunctions of
the narrower and wider slits, respectively. In Figs. 2(a), (c), (e)
and (f) the absolute values are plotted because these quantities are
complex. The eigenmodes in the narrower and wider slits are denoted
by $\psi_{n}^{(na)}$ and $\psi_{n}^{(wi)}$, respectively. In Figs.
2(e) and (f), the lowest modes $\psi_{1}^{(na)}$ and
$\psi_{1}^{(wi)}$, plotted by the black curves, are just SPP modes,
and the second modes $\psi_{2}^{(na)}$ and $\psi_{2}^{(wi)}$,
plotted by the red curves, are of actually antisymmetric wave
functions within the slits. Note that the curve of
$|\psi_{3}^{(wi)}|$ is divided into three parts by two zeros. The
sign of the central part of $\psi_{3}^{(wi)}$ is contrary to the
other parts.

We notice that under our present parameters, only the first two
eigenmodes of the narrower slit $\psi_{1}^{(na)}$ and
$\psi_{2}^{(na)}$ are within the slit (guided modes), see the black
and red lines in Fig. 2(e). The wave functions of the higher modes
mainly distribute within the metal (radiation modes), see, as an
example, the third mode $\psi_{3}^{(na)}$ in Fig. 2(e). The behavior
of $|\psi_{3}^{(na)}|$ with the position of the narrower slit has to
be explicitly given as following. When the slit moves rightwards,
the width of metal at the right side of the slit becomes thinner, so
that the hill is compressed. If the narrower slit is on the right
side of the center position $x=1$ $\mu$m, the hill will appear at
the left side of the slit. While if the slit is just at or very near
the center $x=1$ $\mu$m, there will be two hills at the two sides of
the slit, respectively, since the structure in this case is
symmetric [6].

Three obvious features of excitation efficiencies can be seen in
Figs. 2(a) and (c). The first is that all the curves there exhibit a
central symmetry, because all the configurations are symmetric with
respect to the central line at $x=1$ $\mu$m. The second is, from
comparison of the black and red solid lines in Figs. 2(a) and (c),
that the excitation efficiencies of the SPP modes in narrow slits
are much larger than those of the second modes. The third is, by
inspection of dash-dotted lines in Figs. 2(a) and (c), that the
shapes of the efficiency curves of the modes in wider slits resemble
their eigenfunctions $|\psi_{n}^{(wi)}|$ in Fig. 2(f). The latter
two features can be attributed to the treatment of MEM which
involves a mutual expansion between the modes in different layers.

We should keep in mind that the total field at the layer boundary,
$H_{z}(x)$, can be respectively achieved by the linear combination
of eigenfunctions in the narrower and wider slit, and the expansion
coefficients depend on the position of the narrower slit, subject to
boundary conditions. Then, the curves in Figs. 2(a) and (c) can be
explained qualitatively.

We first see the case where the wave is incident from the narrower
slit to wider one, as shown by the inset in Fig. 2(a). For reflected
waves, the reflection efficiencies $|R_{n}|$ are proportional to the
projection $\int_{0}^{L}\psi_{n}^{(na)}(x)H_{z}(x)dx$ where the
integration is along the interface between the narrower and wider
slits. Since the eigenfunctions in the wider slit or their
combination, $H_{z}(x)$, can be seen as a smooth variation within
the range of narrower slit, the reflection efficiency of the SPP
mode in the narrower slit,
$|R_{1}|\propto\int_{0}^{L}\psi_{1}^{(na)}(x)H_{z}(x)dx\approx\int_{w^{(na)}}\psi_{1}^{(na)}(x)H_{z}(x)dx$,
is dominant because its eigenfunction is also smooth within the
slit, and that
$|R_{2}|\propto\int_{0}^{L}\psi_{2}^{(na)}(x)H_{z}(x)dx\approx\int_{w^{(na)}}\psi_{2}^{(na)}(x)H_{z}(x)dx$
is very small because the second mode is an antisymmetric function
within the slit. For the transmitted waves, the transmission
efficiencies $|T_{n}|$ are qualitatively determined by
$\int_{0}^{L}\psi_{n}^{(wi)}(x)H_{z}(x)dx$, which includes
$\int_{0}^{L}\psi_{n}^{(wi)}(x)\psi_{1}^{(na)}(x)dx$,
$\int_{0}^{L}\psi_{n}^{(wi)}(x)\psi_{2}^{(na)}(x)dx$,
$\int_{0}^{L}\psi_{n}^{(wi)}(x)\psi_{3}^{(na)}(x)dx$, and so on. We
have already known that the excitation of the second mode in the
narrow slit is very small, so that the contribution of the factor
$\int_{0}^{L}\psi_{n}^{(wi)}(x)\psi_{2}^{(na)}(x)dx$ is negligible.
The contribution of the radiation modes is relatively complicated,
but unimportant because what happened inside the slit is the key
part of the scattering procedure; while the radiation modes
localized in metal are excited to fulfill the boundary condition
outside the slit. That is why we try to avoid theses modes in the
discussion. By several numerical tests, it is sure that the
radiation modes do have contribution to the transmitted waves but
the contribution is comparatively small. Therefore, the factor
$\int_{0}^{L}\psi_{n}^{(wi)}(x)\psi_{1}^{(na)}(x)dx\approx\int_{w^{(na)}}\psi_{n}^{(wi)}(x)\psi_{1}^{(na)}(x)dx$
mainly determines the transmission efficiencies $|T_{n}|$. As an
example, let us see the $|T_{3}|$ curve.
$|T_{3}|\propto\int_{w^{(na)}}\psi_{3}^{(wi)}(x)\psi_{1}^{(na)}(x)dx$,
where $\psi_{1}^{(na)}$ is smooth within a narrow region, see, the
black line in Fig. 2. When the narrower slit is positioned at the
left side with its center being at $x=0.65$ $\mu$m,
$|\psi_{3}^{(wi)}|$ has a maximum at this position. Therefore the
projection of $\psi_{1}^{(na)}$ onto $\psi_{3}^{(wi)}$ is at a
maximum. As the narrower slit moves rightwards, we image that the
black curve in Fig. 2(e) shifts rightwards. At $x=0.794$ $\mu$m,
$|\psi_{3}^{(wi)}|$ is zero. Accordingly, the projection of
$\psi_{1}^{(na)}$ at this position onto $\psi_{3}^{(wi)}$, as well
as $|T_{3}|$, reaches zero. Between $x=0.65$ and $x=0.794$ $\mu$m,
$|T_{3}|$ should drop from the maximum to zero. We notice that
around the zero, the phase of $T_{3}$ changes nearly $\pi$. At the
other zero of $|\psi_{3}^{(wi)}|$ at $x=1.206$ $\mu$m, $|T_{3}|$
again reaches zero and its phase changes nearly $\pi$ once more.
This analysis explains why the shape of $|T_{3}|$ is like to
$|\psi_{3}^{(wi)}|$. It is the narrow and smooth profile of
$|\psi_{1}^{(na)}|$ that causes the similarity of the curves between
$|T_{3}|$ and $|\psi_{3}^{(wi)}|$ curves. The $|T_{2}|$ curve in
Fig. 2(a) is understood in the same way. $|T_{1}|$ is mainly
determined by
$\int_{w^{(na)}}\psi_{1}^{(wi)}(x)\psi_{1}^{(na)}(x)dx$, which is a
smooth and relatively flat curve due to the smooth variations of
both SPP waves.

We next see the case where a SPP wave is incident from the wider
slit to narrower one, as shown in the inset of Fig. 2(c), the
incident wave being a smooth curve within the range of the wider
slit width, see the black curve in Fig. 2(f). We again begin with
the waves in narrower slit. $|T_{n}|$ is proportional to the
integral $\int_{0}^{L}\psi_{n}^{(na)}(x)H_{z}(x)dx$, where
$H_{z}(x)$ is the combination of eigenfunctions in the wider slit
and considered as a smooth varying curve within the range of the
narrower slit, so that the transmission efficiency of the SPP mode
is dominant and much larger than that of the second mode. For the
waves in the wilder slit, ignoring the contribution of the second
mode and radiation modes, the reflection efficiencies $|R_{n}|$ is
mainly determined by
$\int_{w^{(na)}}\psi_{n}^{(wi)}(x)\psi_{1}^{(na)}(x)dx$, leading to
the fact that the $|R_{n}|$ curves in Fig. 2(c) have similar shapes
as $|T_{n}|$ curves in Fig. 2(a).

The transmission efficiency $|T_{1}|$ in Fig. 2(a) is exactly the
same as that in Fig. 2(c), and the efficiencies $|R_{1}|$, $|T_{2}|$
and $|T_{3}|$ in Fig. 2(a) have the same behavior as $|R_{1}|$,
$|R_{2}|$ and $|R_{3}|$ in Fig. 2(c), respectively, although with
different values. The reflection efficiencies $|R_{1}|$ and
$|R_{2}|$ in Fig. 2(c) are higher than those in Fig. 2(a). This is
because Fig. 2(c) represents the case that wave incident from a
wider slit to a narrower one, which needs to squeeze light into a
narrower space, so the higher reflection is understandable.

The variations of the scattering phase shifts for the above two
different incident cases are plotted in Fig. 2(b) and (d). It is
seen that the phases of both $T_{1}$ in these two figures are also
the same. We note that at the positions where $\psi_{n}^{(wi)}$ is
zero, the corresponding coefficients $R_{n}$ and $T_{n}$ have phase
change of $\pi$.

The results of Type II structure are plotted in Fig. 3. The left
walls of the slits are aligned at $x_{1}=0.6$ $\mu$m. The width of
the narrower slit is $0.1$ $\mu$m, but that of the wider one,
denoted by $w$, varies from $0.1$ to $0.8$ $\mu$m, as shown in the
insets in Figs. 3(a) and (c). The most distinct feature is the
drastic changes of the excitation efficiencies over a narrow range
of slit width, as shown in Figs. 3(a) and (c) near $w=0.46$ $\mu$m.

\begin{figure}[htbp]
\centering\includegraphics  [width=6cm] {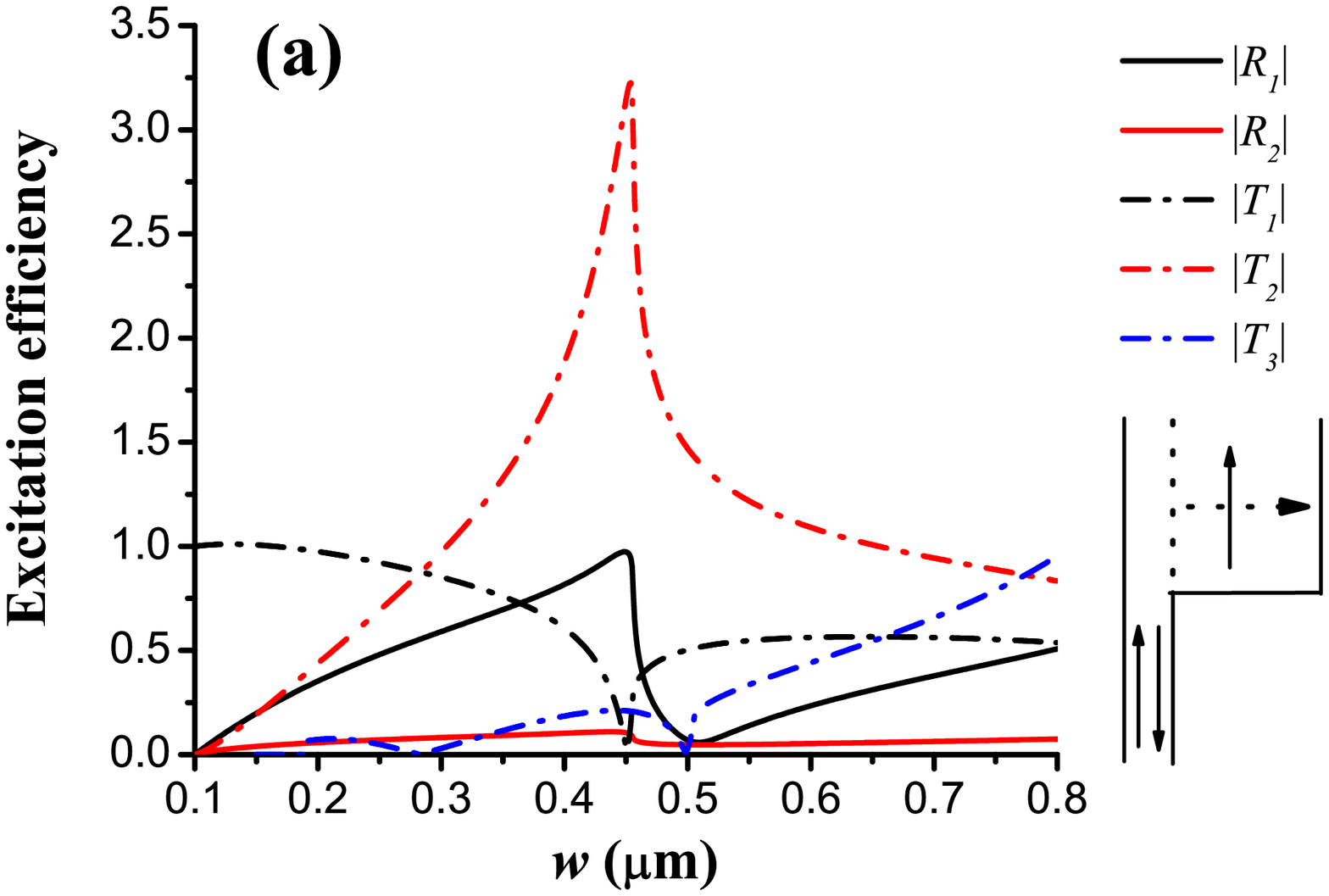}
\includegraphics  [width=6cm] {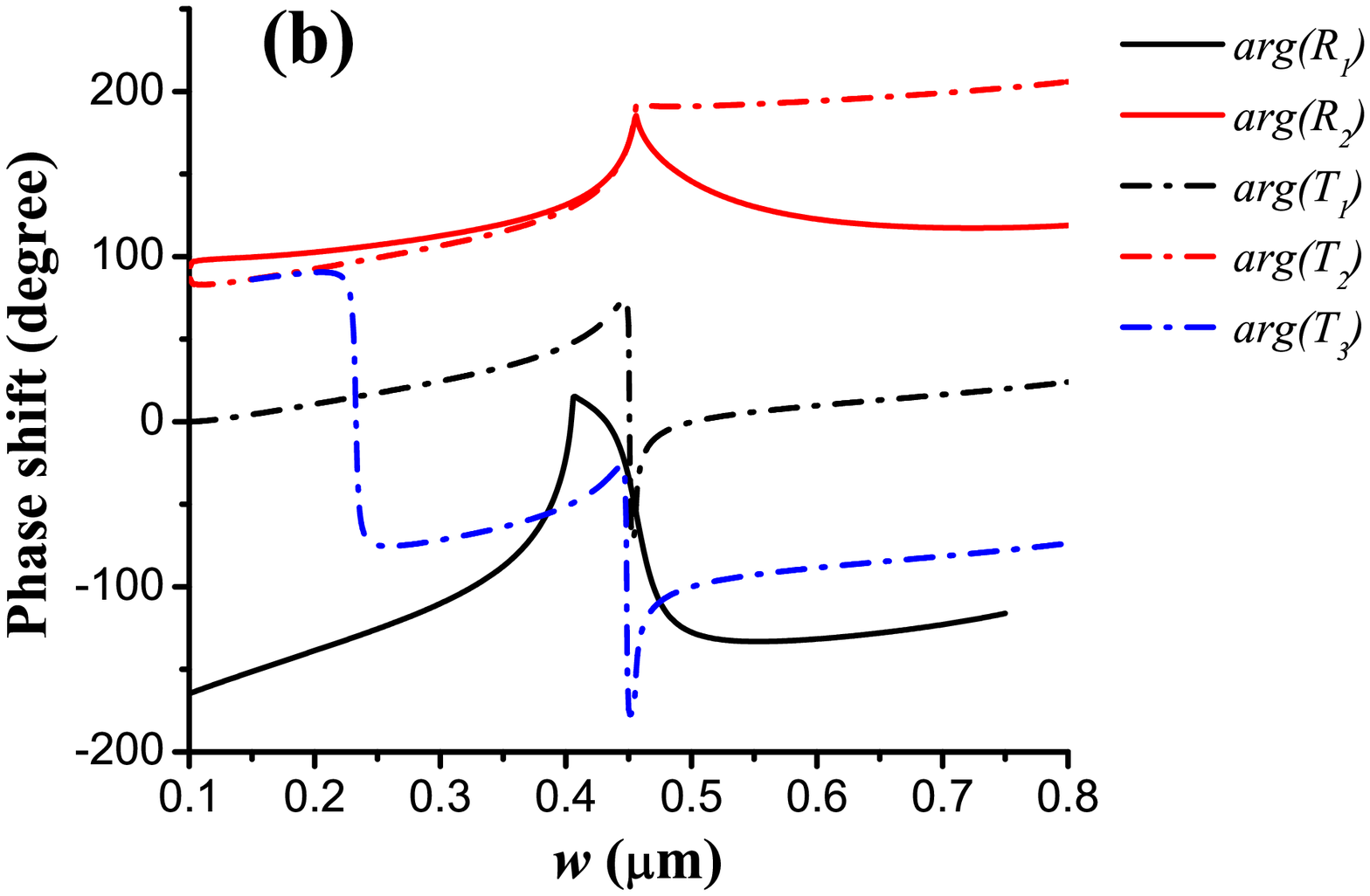}
\includegraphics  [width=6cm] {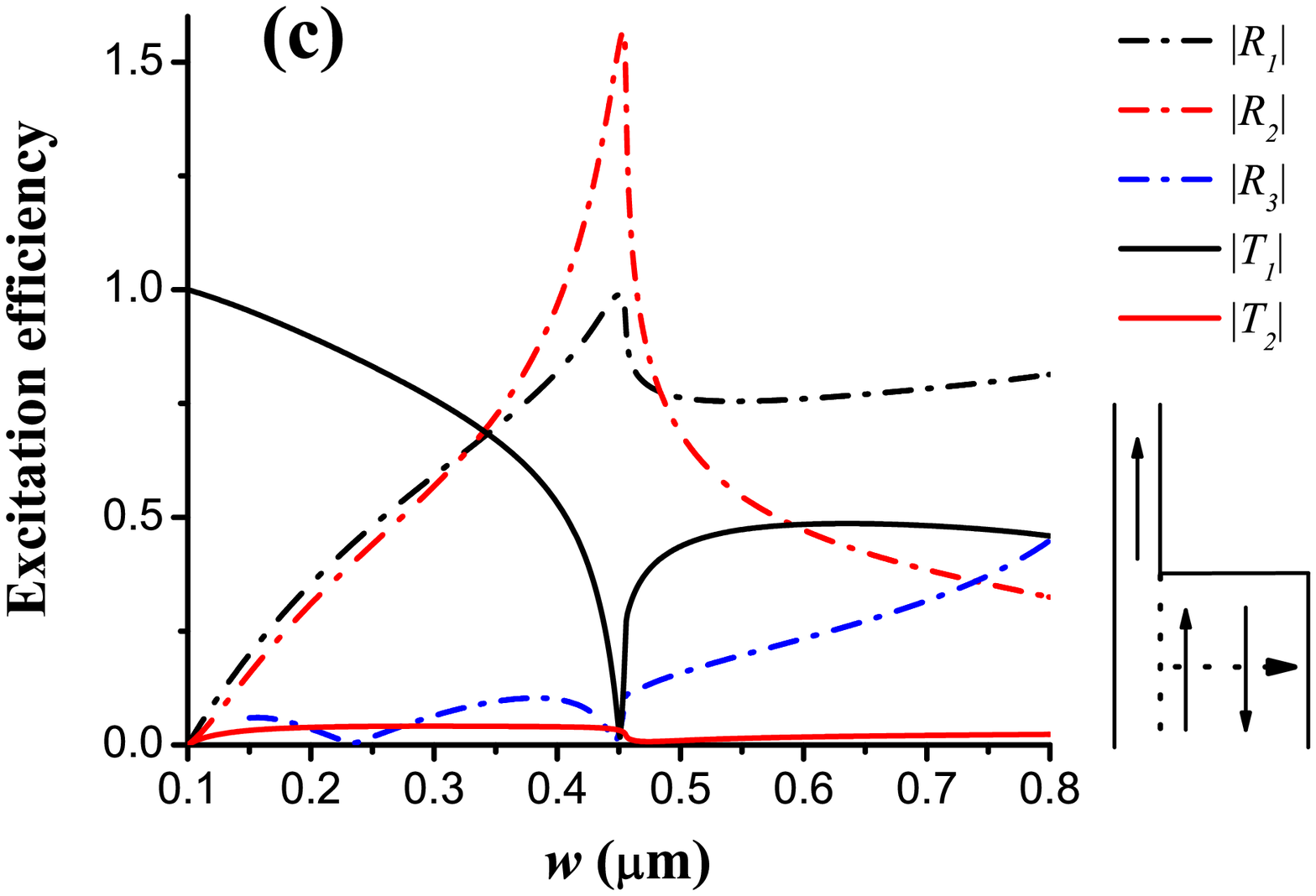}
\includegraphics  [width=6cm] {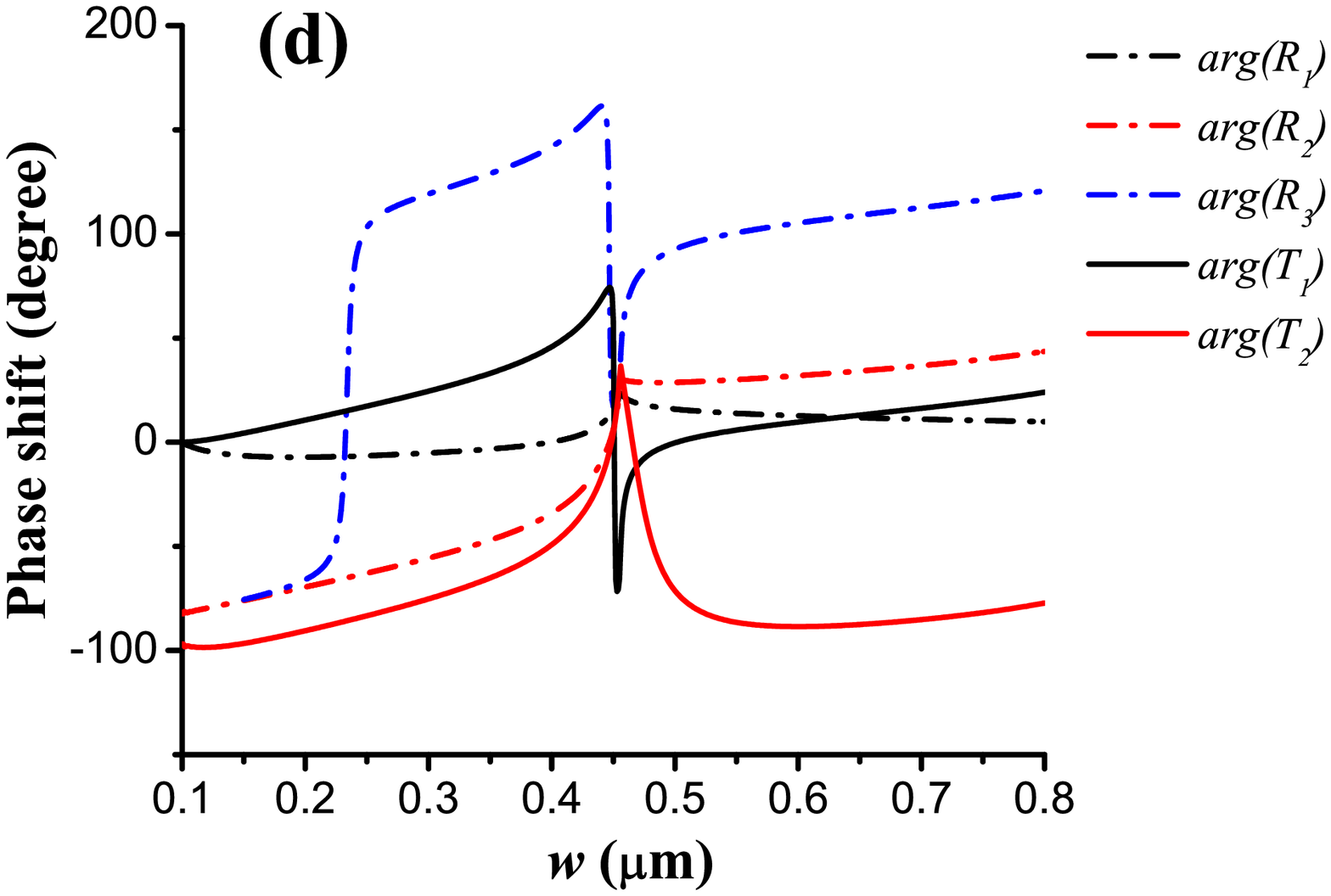}
\includegraphics  [width=6cm] {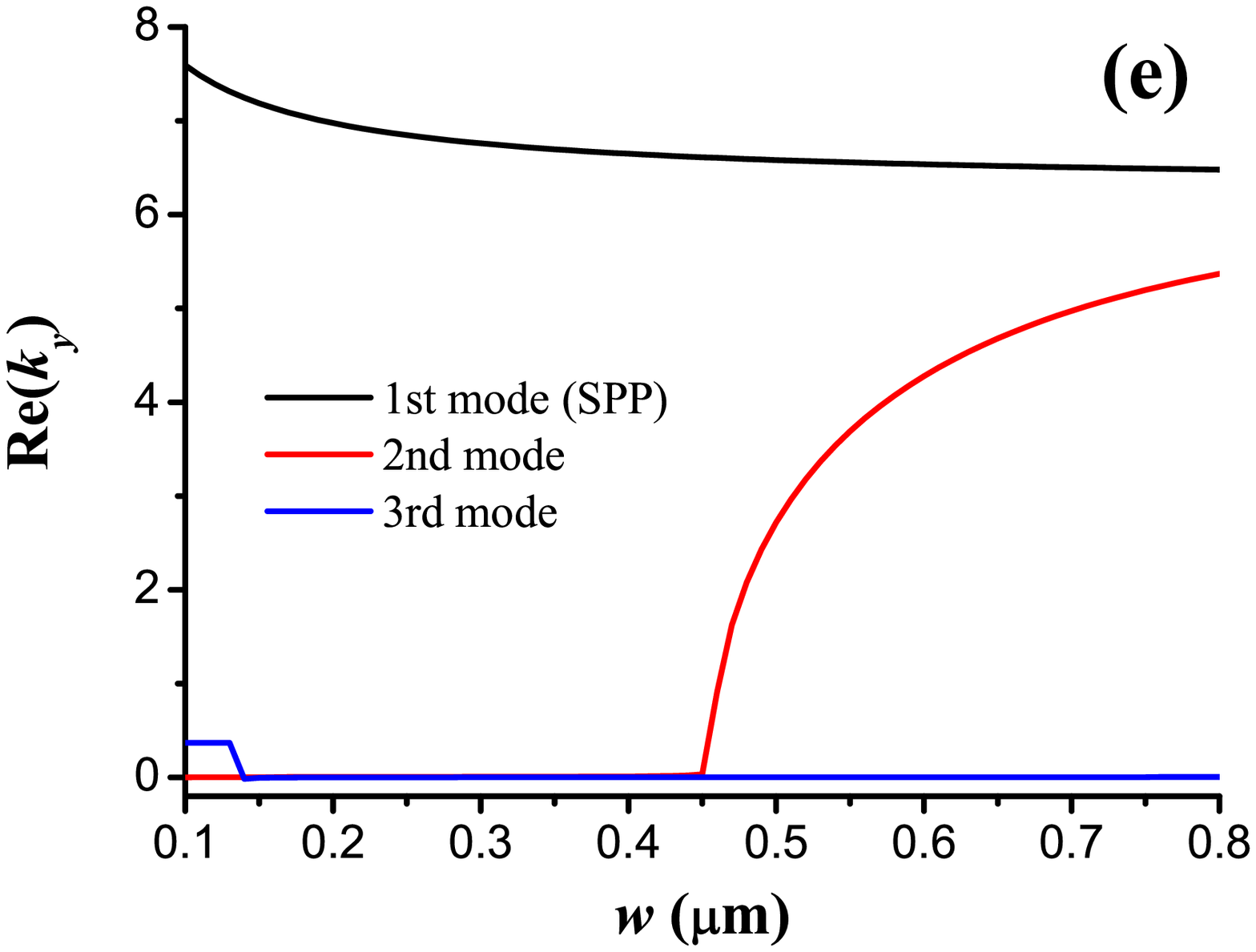}
\includegraphics  [width=6cm] {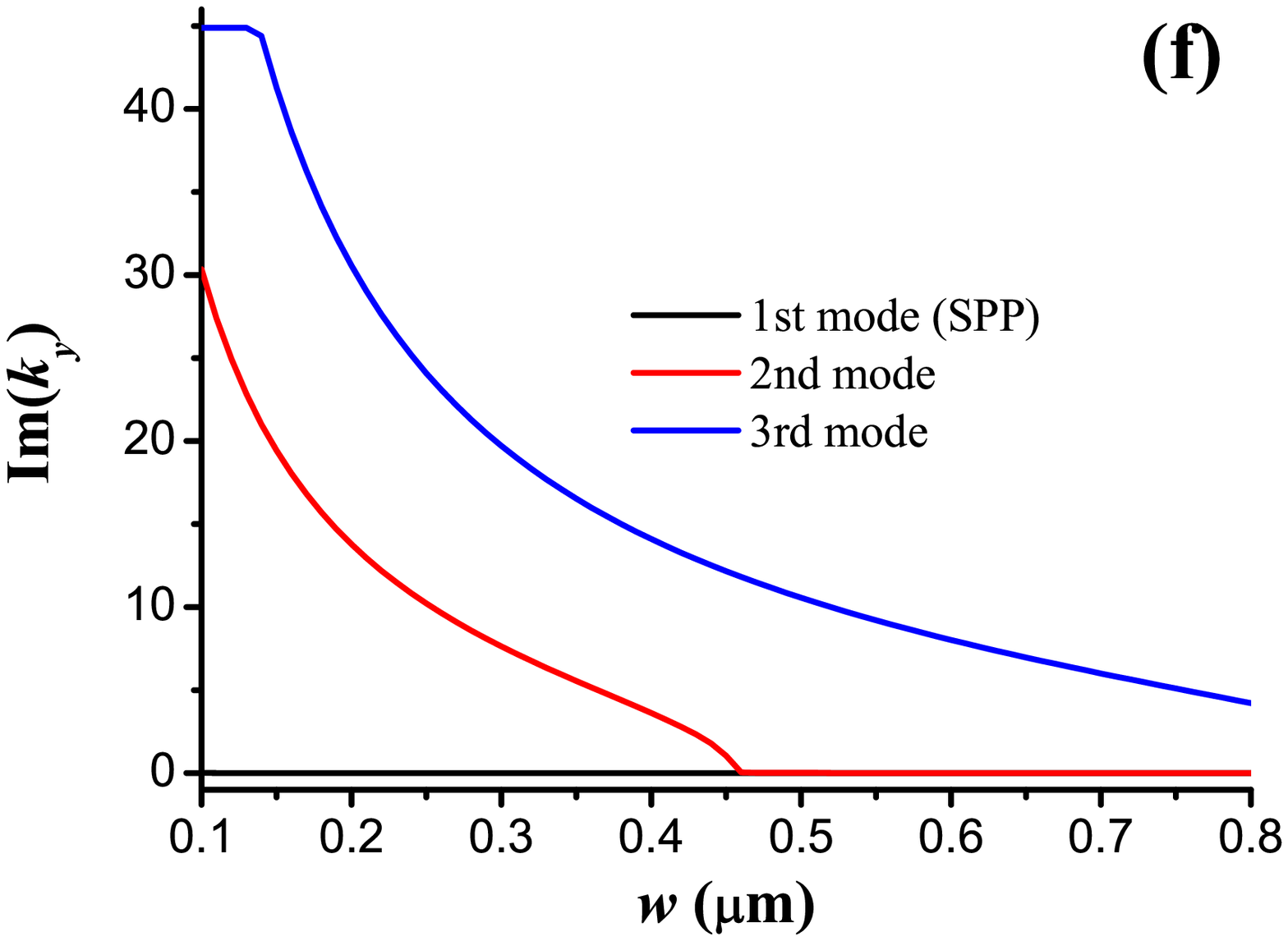}
 \caption{(color online). Scattering in Type II junction structures. The left
walls of the slits are aligned at $x_{1}=0.6$ $\mu$m.
$Q^{(0)}=Q^{(1)}=Q^{(2)}=0$. In all the figures, the results of the
wider slit are plotted by solid lines and those of the narrower slit
by dash-dotted lines. The width of the narrower slit is $0.1$
$\mu$m. The $x$-axes are the width of the wider slit, denoted as
$w$. For wave incident from the narrower slit to the wider one, (a)
excitation efficiency; (b) phase shift. For wave incident from the
wider slit to the narrower one, (c) excitation efficiency; (d) phase
shift. In (e) and (f) plotted are the real and imaginary parts of
the propagation constant $k_{y}$, appearing in Eq. (1) with the
notation $k_{y}^{(l)}$, of the wider slit.}
\end{figure}

When the width of a slit is $0.1$ and $0.8$ $\mu$m, the first three
eigenmodes have been plotted in Figs. 2(e) and (f), respectively.
Now the width $w$ varies. Our calculation shows that the modes
resemble those in Fig. 2(e) when $w<0.15$ $\mu$m and those in Fig.
2(f) otherwise. The first three eigenmodes are here labeled by the
notations $\psi_{1}$, $\psi_{2}$ and $\psi_{3}$, and their $k_{y}$'s
are by $k_{y1}$, $k_{y2}$ and $k_{y3}$, respectively.

The SPP mode $\psi_{1}$ is obviously a propagation one since
$k_{y1}$ has a negligible imaginary part, and $\psi_{3}$ is a
decaying one as demonstrated by the large imaginary part of
$k_{y3}$. When $w<0.46$ $\mu$m, $k_{y2}$ is nearly purely imaginary
so that $\psi_{2}$ is an evanescent mode, while when $w>0.46$
$\mu$m, $k_{y2}$ becomes nearly purely real so that $\psi_{2}$ turns
to be a propagation mode. A turning point appears at $w=0.46$ $\mu$m
at which $\psi_{2}$ transforms its propagation property. This
transformation leads to the drastic changes of the excitation
efficiencies, similar to the cause of the well-known Wood's anomaly
in the grating theory [21].

The analysis about the excitation efficiencies in Figs. 3(a) and (c)
are in the same way as those in the Type I structure. Therefore,
some similar conclusions are obtained, such as the identity of
$|T_{1}|$ curves in Figs. 3(a) and (c) and the similarity between
the efficiencies $|R_{1}|$, $|T_{2}|$ and $|T_{3}|$ in Fig. 3(a) and
$|R_{1}|$, $|R_{2}|$ and $|R_{3}|$ in Fig. 3(c), respectively,
although with different values.

The drastic changes of excitation efficiencies at the turning point
have to be explained from an energy perspective. As an example, let
us see the case where the wave is incident from the narrower slit to
wider one, as shown by the inset in Fig. 3(a). At the start point
$w=0.1$ $\mu$m, the two slits are the same, so that $|T_{1}|=1$ and
all other excitation efficiencies are zero. Close to the turning
point, $|T_{1}|$ reaches the minimum, and $|R_{1}|$ and $|T_{2}|$
reach the maximum. Since the second mode in the wider slit
$\psi_{2}$ now is an evanescent mode, the energy is mainly stored in
the reflected SPP wave. Once $\psi_{2}$ becomes a propagation mode,
it must gain a large portion of energy from the reflection due to
its large excitation efficiency, and leads to the rapid drop of
$|R_{1}|$ as shown in Fig. 3(a). At the same time, the dropping
$|R_{1}|$ further causes a redistribution of excitation efficiencies
by the boundary continuum condition. Therefore, what behind the
drastic changes of excitation efficiencies is a redistribution of
energy between evanescent modes and propagation modes. The
explanation of the energy redistribution is also suitable for the
case where the wave is incident from the wider slit to narrower one.

It is seen from Figs. 3(b) and (d) that the phases of both $T_{1}$
in these two figures are also the same. At the turning point,
$T_{1}$ changes its phase by $\pi$. The change of $\pi$ in phase at
the turning point also occurs for $T_{3}$ in Figs. 3(b) and $R_{3}$
in Figs. 3(d).

So far, the scattering mechanisms of the two types of structures are
investigated. Although the investigation here is restricted to the
incident wave with wavelength $\lambda_{0}=1$ $\mu$m only, the
analysis above is also applicable to the infrared and visible
spectrum. Furthermore, the analysis is important in practical
application. For example, one can excite/suppress specific modes to
control the field distribution inside a slit, or design high
efficient reflector, by changing the position or width of the slit.

\subsection {The multi-mode multi-reflection model}

Having had the knowledge of the scattering of the interface in a
single junction structure, we are ready in this subsection to
discuss the transmission in a step-modulated slit which can be
regarded as the combination of two junction structures. In order to
reveal the transmission clearly, we present here an analysis of
multi-mode multi-reflection model that combines wave and ray optics.

Figure 4 is the sketch of the multi-mode multi-reflection model. In
a step-modulated subwavelength metal slit, there are two interfaces
at $Q^{(1)}$ and $Q^{(2)}$. Let us discuss the wave reflection and
transmission in the slit. When the incident SPP launched from
$Q^{(0)}$ in Layer 1 impinges the interface between Layers 1 and 2,
$Q^{(1)}$, it generates the reflection wave in Layer 1 and
transmission wave in Layer 2. The latter continues going upwards,
and when reaching other interface $Q^{(2)}$, yields reflection and
transmission waves again. Obviously, there occurs multi-reflection
in Layer 2, shown by Fig. 4(a).

\begin{figure}[htbp]
\centering\includegraphics  [width=6.5cm] {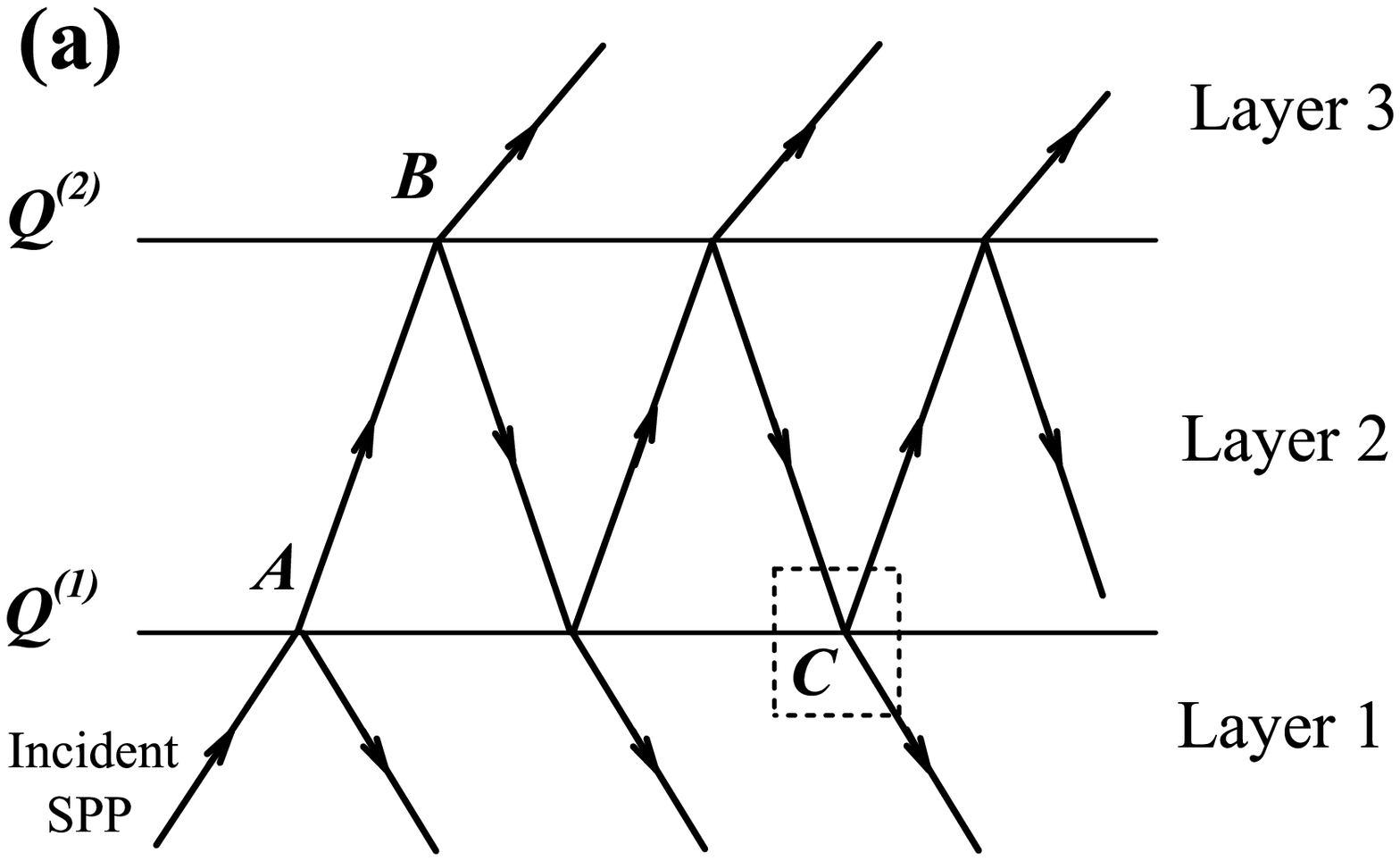}
\includegraphics  [width=6.5cm] {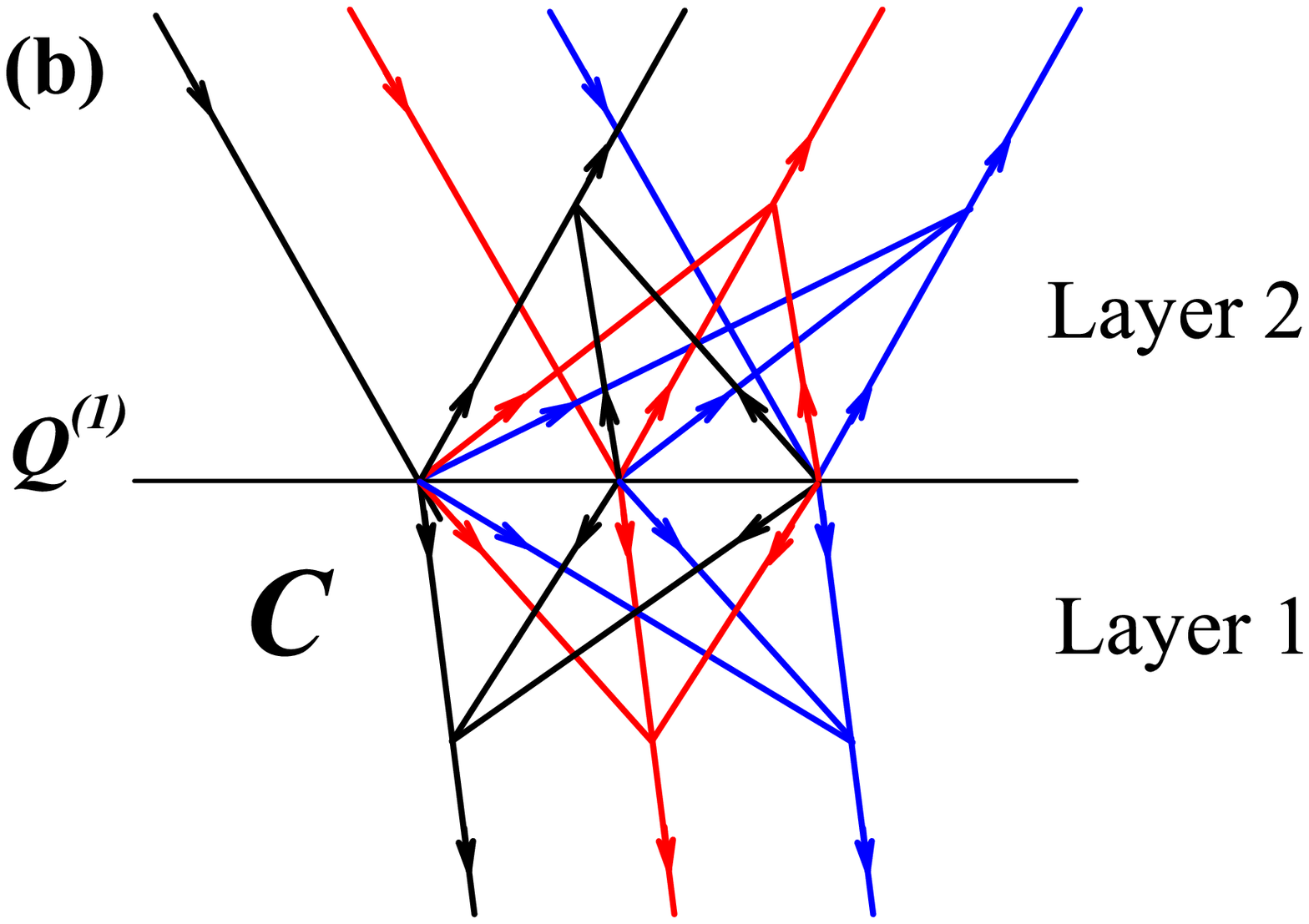}
 \caption{(color online). Sketch of the multi-mode multi-reflection model. (a)
multi-reflection between interfaces and (b) multi-mode excitations
at each point.}
\end{figure}

Because the wave in each layer is the linear combination of the
eigenmodes of the layer, each ray in Fig. 4(a) can in fact be
expanded by eigenmodes, except the primary incident light. For
example, at the point A, the reflected wave contains all possible
modes in Layer 1 and the transmitted wave contains the modes in
Layer 2. That is to say, the scattering excites all the modes in
both layers. When all the possible modes in Layer 2 reach point B,
each mode again excites all possible eigenmodes in reflected wave in
Layer 2 and in transmitted wave in Layer 3. The phenomenon is termed
as multi-mode excitation. To show the phenomenon explicitly, we draw
in Fig. 4(b) the multi-mode excitation at point C. Suppose that the
waves in Layers 1 and 2 are expanded by three eigenmodes,
respectively. Then when the three modes in Layer 2 are incident to
point C, as shown in Fig. 4(b), the first mode yields the reflected
and transmitted waves, both containing three eigenmodes in
respective layer, i.e., the incident black line excites the black,
red and blue lines in the transmitted waves in Layer 1 and reflected
waves in Layer 2, respectively. In the same way, the incident red
line also excites the black, red and blue lines in the transmitted
waves in Layers 1 and reflected waves in Layer 2, respectively, and
so does the incident blue line. Therefore, the total reflected wave
at point C includes three eigenmodes in Layer 2, each being in turn
the superposition of the reflections from the three incident
eigenmodes. Similarly, the total transmitted wave at point C
includes three eigenmodes in Layer 1, each being in turn the
superposition of the transmissions from the three incident
eigenmodes.

In summary, the total transmission and reflection coefficients in
Layer 3 and Layer 1 in Fig. 4(a) are obtained by summing up all the
single-scattered coefficients, respectively. In addition, it is
worth mentioning that the multi-mode multi-reflection model is a
generalized form of the single-mode multi-reflection model which
occurs in a F-P cavity. The former will be simplified to be the
latter if only one mode can be excited.

The physical explanation of the multi-mode multi-reflection process
is named as model analysis. In order to testify this analysis,
numerical calculation based on this physical picture is carried out
and the results are compared with MEM. In Fig. 5 plotted are the
transmission efficiencies as a function of the length of Layer 2
$q^{(2)}$ when the incident wave is SPP mode. The solid lines in
Fig. 5(a) and (b) are the results of MEM, which surely comprise the
contributions from all possible eigenmodes. The symbols are the
results from the model analysis. In a slit with width $w^{(2)}=0.3$
$\mu$m, only the first mode, i.e., the SPP mode, can propagate and
all the other modes are evanescent. Thus, when $q^{(2)}$ is
sufficiently long, the higher modes attenuate to a negligible value,
and the transmission can be well described by a multi-reflection of
only the SPP mode.

\begin{figure}[htbp]
\centering\includegraphics  [width=6.5cm] {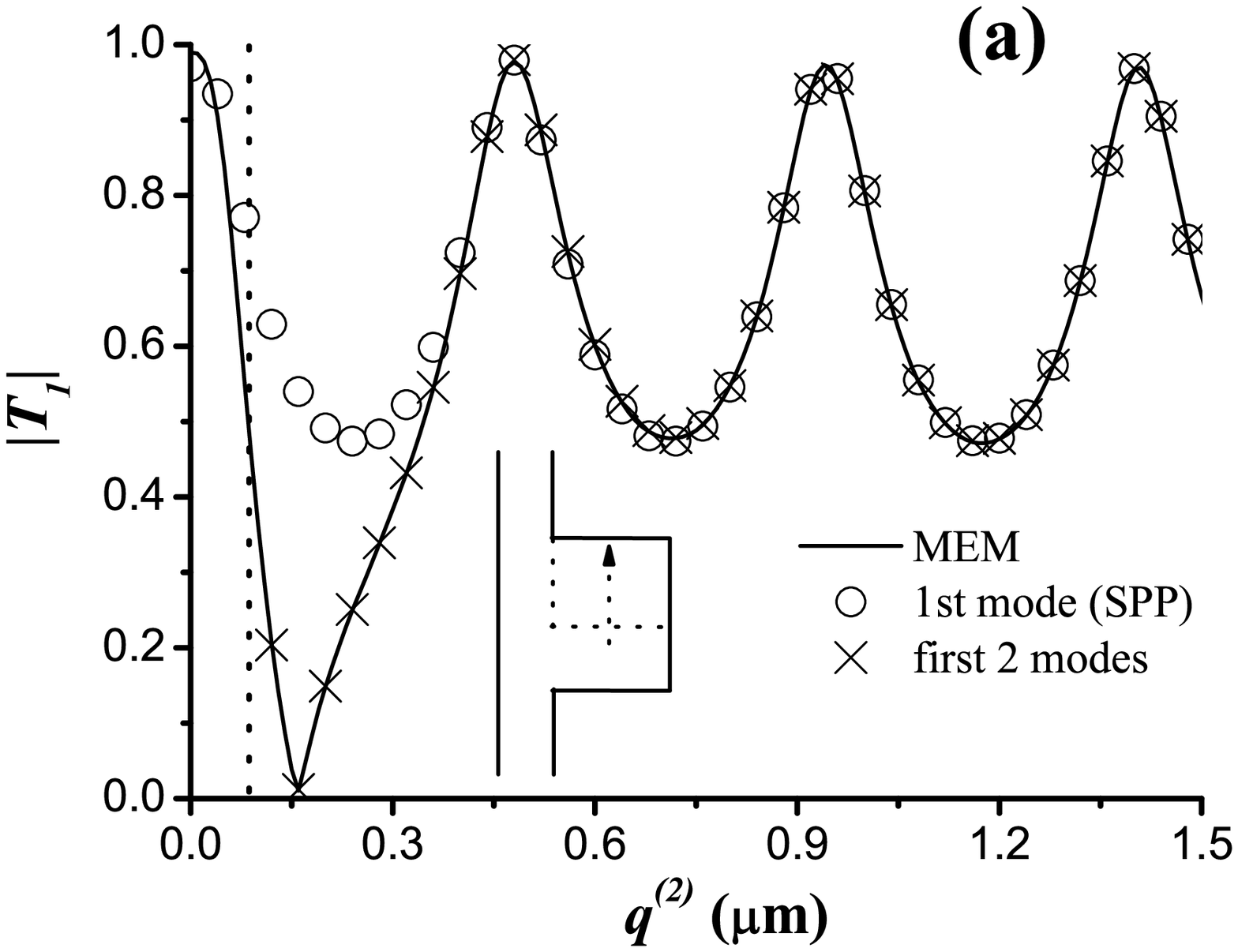}
\includegraphics  [width=6.5cm] {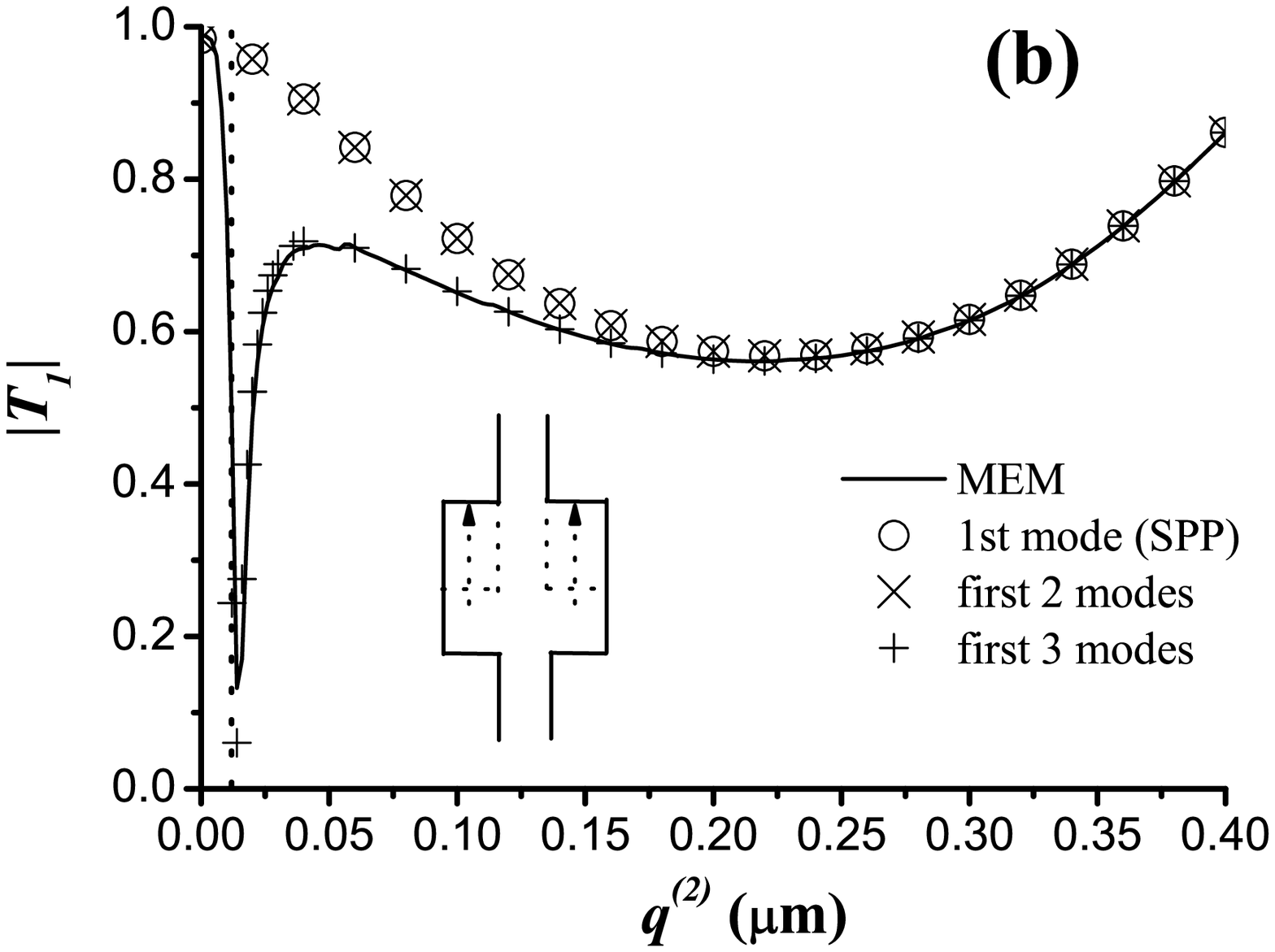}
 \caption{The SPP transmission vs. $q^{(2)}$ calculated by MEM and model
analysis for slit structure with
$\left[Q^{(0)},Q^{(1)},Q^{(2)}\right]=\left[-1,0,q^{(2)}\right]$
$\mu$m. The solid lines are the results of MEM. The circles, crosses
and plus signs are the results including contributions from the
first one, two and three modes, respectively, from the model
analysis. (a) Slits align to left at $x_{1}^{(l)}=0.85$ $\mu$m,
$\left[w^{(1)},w^{(2)},w^{(3)}\right]=\left[0.1,0.3,0.1\right]$
$\mu$m; (b) The symmetric case of the slit structure in (a).}
\end{figure}

In Fig. 5(a) it is seen that the results of the model analysis
including the first two modes are accurately the same as the line
from MEM. When $q^{(2)}>0.8$ $\mu$m, the circles and crosses are
identical, indicating that the contribution from the second mode is
negligible. While for $q^{(2)}<0.8$ $\mu$m, crosses deviate from
circles, indicating that the second mode should not be omitted since
it does not fade out within this distance range. In Fig. 5(a), the
crosses end at $q^{(2)}=0.09$ $\mu$m, because below this distance
the multi-reflection of the first two modes diverges. What is the
reason of the divergence? Firstly, the divergence is not caused by
the propagation mode since the multi-reflection of the SPP mode
always converges, as shown by the circles in Fig. 5(a). Secondly, it
is neither caused by the exponentially increasing term which
originates from improperly handling the evanescent waves [20] for it
occurs only at short distances. Actually, the divergence arises from
the coupling between the eigenmodes containing the evanescent modes.
With the contribution of the evanescent wave, as shown in Fig. 4(b),
the superposition will result in a larger transmission and
reflection coefficients after each scattering if the second mode
does not decay to a certain value. Thus, there exists a critical
distance above which the multi-mode multi-reflection analysis is
applicable. For the structure given in Fig. 5(a), it is
$q^{(2)}=0.09$ $\mu$m.

To verify the above conclusion about the divergence of the
evanescent mode, we suppress the antisymmetric second mode by
reforming the slit structure to a symmetric one, as shown in Fig.
5(b). Then the circles and crosses are identical at any distance.
Let us see the contribution from the third mode. The plus signs
including contributions from the first three modes are in good
agreement with the results of MEM as $q^{(2)}>0.012$ $\mu$m. The
crosses and circles are identical when $q^{(2)}$ is above $0.3$
$\mu$m, but it is not so when $q^{(2)}$ is below $0.3$ $\mu$m. That
is to say, if the distance is less than $0.3$ $\mu$m, the evanescent
third mode is not negligible. This time the critical height for the
third mode is $q^{(2)}=0.012$ $\mu$m, which is much smaller than
that of the second mode in the structure shown in Fig. 5(a). The
reason is that the decay of the third mode is faster than that of
the second mode, for the propagation constant $k_{y}$ of the former
has a larger imaginary part than the latter, as shown in Fig. 3(f).

\begin{figure}[htbp]
\centering\includegraphics  [width=6.5cm] {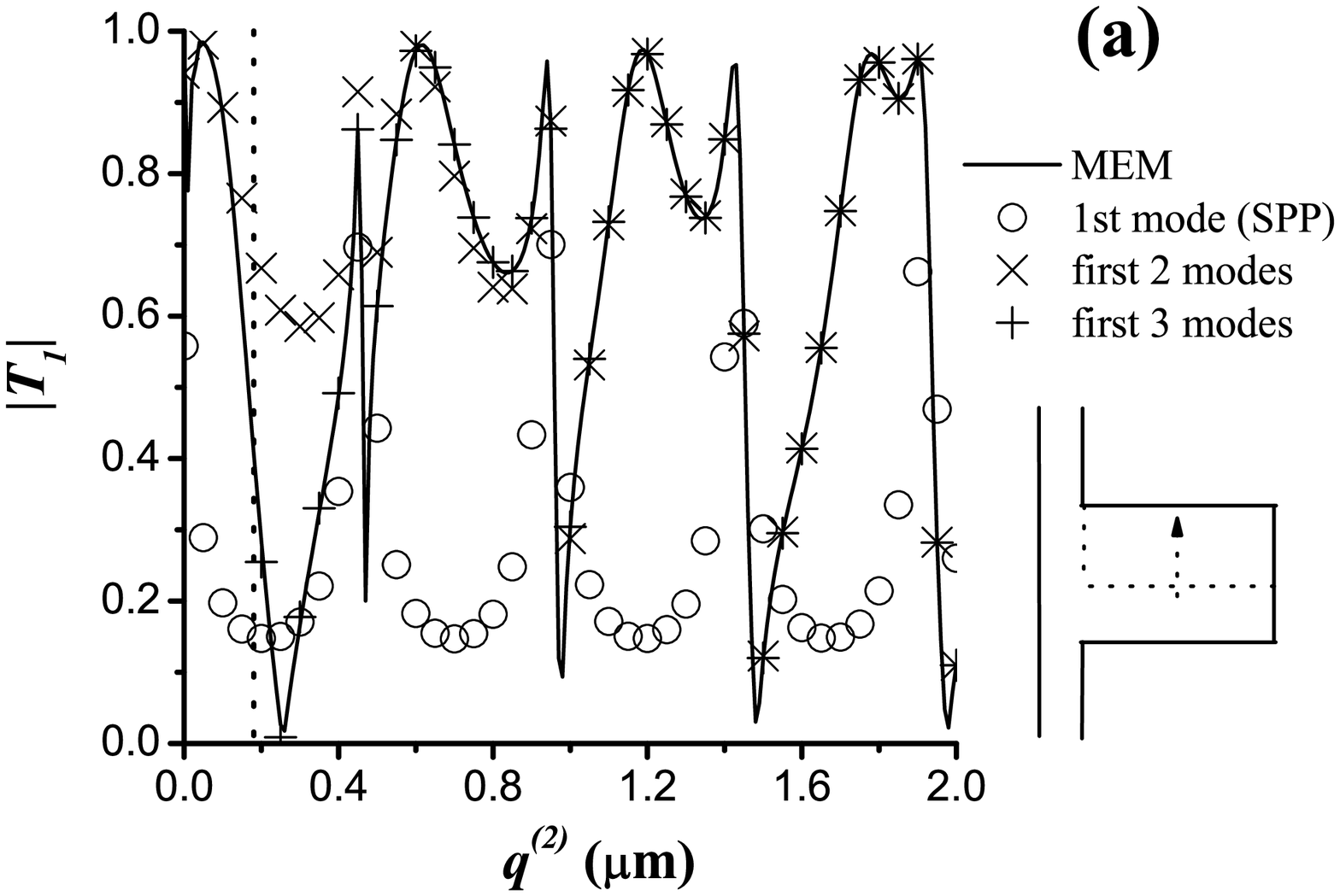}
\includegraphics  [width=6.5cm] {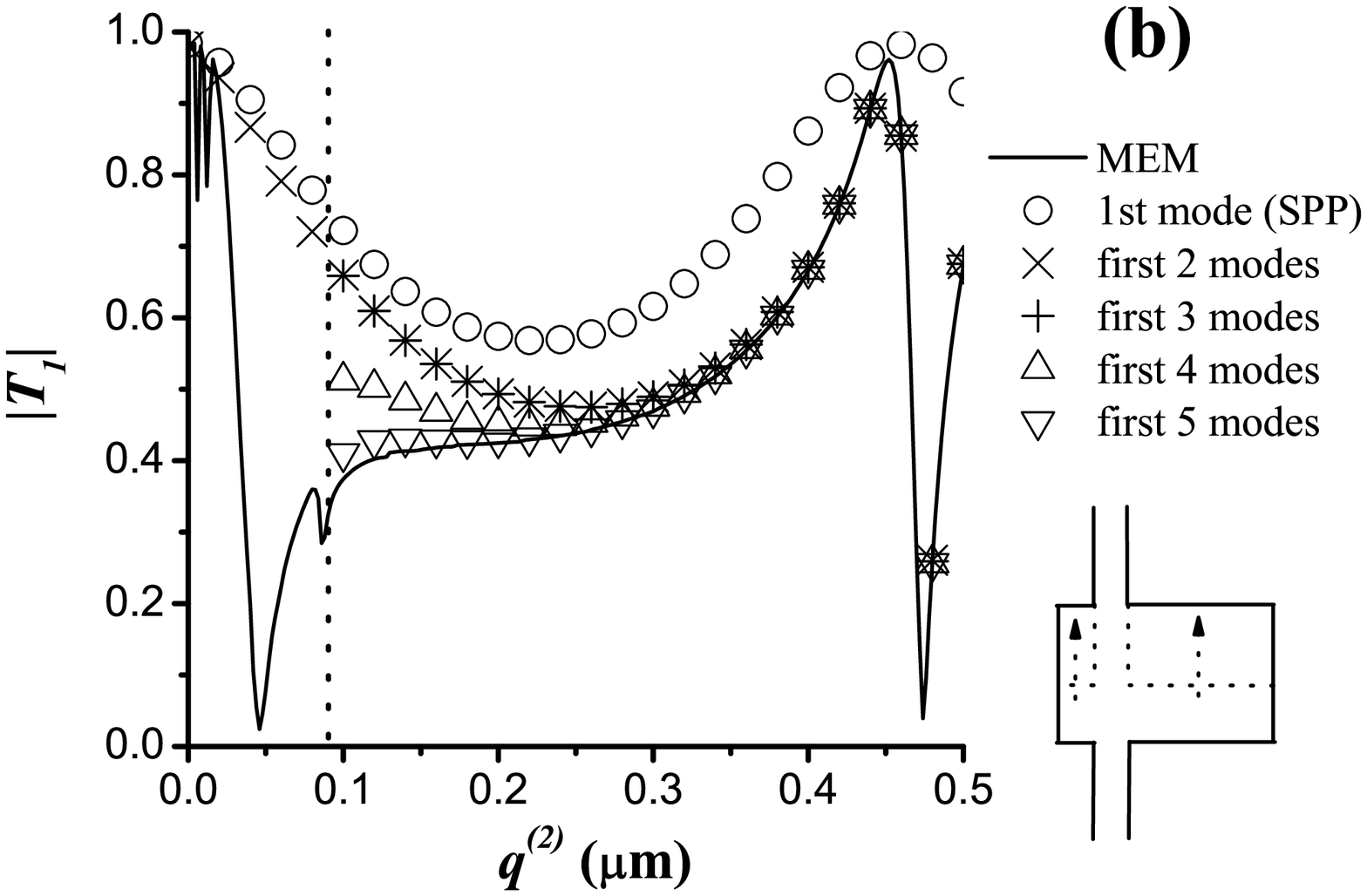}
 \caption{The SPP transmission vs. $q^{(2)}$ calculated by MEM and model
analysis for slit structure with
$\left[Q^{(0)},Q^{(1)},Q^{(2)}\right]=\left[-1,0,q^{(2)}\right]$
$\mu$m. The solid lines are the results of MEM. The circles,
crosses, plus signs, up-triangles and down-triangles are the results
including contributions from the first one to five modes,
respectively, from the model analysis. (a) Slits align to left at
$x_{1}^{(l)}=0.6$ $\mu$m,
$\left[w^{(1)},w^{(2)},w^{(3)}\right]=\left[0.1,0.8,0.1\right]$
$\mu$m; (b) The same structure but the narrower slits are shifted to
be $x_{1}^{(1)}=x_{1}^{(3)}=0.744$ $\mu$m, a position which totally
suppresses the excitation of the 3rd mode, see Fig. 2(a).}
\end{figure}

Next, we investigate the coupling between propagation modes. In Fig.
5(a), only the first mode, the SPP mode, is the propagation one in
Layer 2 with width $w^{(2)}=0.3$ $\mu$m. When the width is enlarged,
the second mode can also become propagating. In Fig. 6(a) plotted
are the transmission efficiencies in the same structures as in Fig.
5(a) except that the width of Layer 2 is extended to be $0.8$
$\mu$m. Under this width, the second mode is indeed propagating. It
is seen that even $q^{(2)}$ gets to zero, the result containing the
contributions from the first two modes is not divergent, and agrees
with the MEM curve very well when $q^{(2)}>1.2$ $\mu$m, which
confirm the statement that the propagation modes do not cause the
divergence. When $q^{(2)}$ is below $1.2$ $\mu$m, the contribution
from the third mode has to be added in order to achieve precise
results. However, as the cost of preciseness, the divergence appears
below $q^{(2)}=0.181$ $\mu$m.

Similar to the treatment in Fig. 5(b) where the second mode in Layer
2 is removed by structural change, it is also possible to suppress
the third mode excited in Layer 2. The way to implement the
suppression is to shift the center of the narrower slits to
$x=0.794$ $\mu$m, as shown in the inset of Fig. 6(b). At this
position, the excitation efficiency of the third mode is nearly
zero, see, Fig. 2(a). The transmission results are plotted in Fig.
6(b). It is seen from the figure that up to the first five
eigenmodes have to be included in the model analysis in order to
meet the MEM curve. The divergence in this case is caused by the
fourth mode, and the corresponding critical width is $q^{(2)}=0.084$
$\mu$m.

In summary, the multi-mode multi-reflection model provides intuitive
and precise description about the transmission inside a
step-modulated subwavelength metal slit when the height of modulated
layer (Layer 2) is above a critical height, while fails below it
because of the coupling between propagation modes and evanescent
modes.

\subsection {Comparison of different methods}

In this subsection, MEM and other three methods, FDTD, SMM, and
ICIM, are discussed, and the calculated results of FDTD and ICIM are
compared to the MEM results. The preciseness of these methods is
investigated and some useful conclusions are obtained. Before
presenting the numerical results, we would like to make a brief
discussion about these four methods.

FDTD, as a commonly used simulation method in optics, is to
calculate field quantities directly from the Maxwell's equations by
difference method. In principle, this method and MEM both can
provide accurate and reliable results. Here we would like to point
out their three discrepancies. Firstly, the way they solve the
Maxwell's equations is different: FDTD uses finite difference method
to evolve fields in space and time domains, while the MEM
establishes and solves the coupled equations in frequency domain by
the method of moments. Secondly, the way they handle outmost
boundaries is different: FDTD makes use of, for the outmost
boundaries of a system, perfectly matched layers which can totally
absorb waves without reflecting them back, while MEM confines the
structure with two perfectly conducting walls such as in this paper.
The feasibility of the latter is due to the fast attenuation of
light (infrared, visible spectrum) in a metal. If the confined width
is large enough, the effects brought by the two perfectly conducting
walls are negligible, as shown in Fig 7(a) below. Although the
perfectly matched layer technique can be introduced to MEM [22], it
dramatically complicates the modal analysis. Thirdly, the way they
converge is different: the convergence of FDTD depends on the size
of the Yee cell used in simulation, while that of MEM on confined
width $L$ and truncation number $N$.

SMM [7-9] and ICIM [13,14] are other two frequently used methods
which show following three features. Firstly, according to the two
methods, the modulated region, Layer 2, would be divided into a
central scattering region and a stub (as shown in the Fig. 2 in Ref.
[7] and Fig. 4 in Ref. [13]), and it would assume that the SPP mode
multi-reflection occurred in the stub (although Refs. [13] and [14]
did not mention this point, it could be recognized from the
transmission equations, Eq. (4) in Ref. [13] and Eq. (8) in Ref.
[14]). Secondly, both of them took the one mode approximation, which
meant that only the SPP modes existed in the stub and slits.
Thirdly, the phase shifts caused by scattering in the central
scattering region could not be calculated properly, so that were
ignored by means of the quasi-statistic approximation [23,13,14]
(although Refs. [7-9] did not mention the quasi-statistic
approximation, it was easily seen by the procedure of obtaining
scattering matrix given in Ref. [8]).

In the following, the numerical comparison between these methods is
performed. The convergence comparison of MEM and FDTD in a Type I
structure is presented in Fig. 7.

\begin{figure}[htbp]
\centering\includegraphics  [width=7cm] {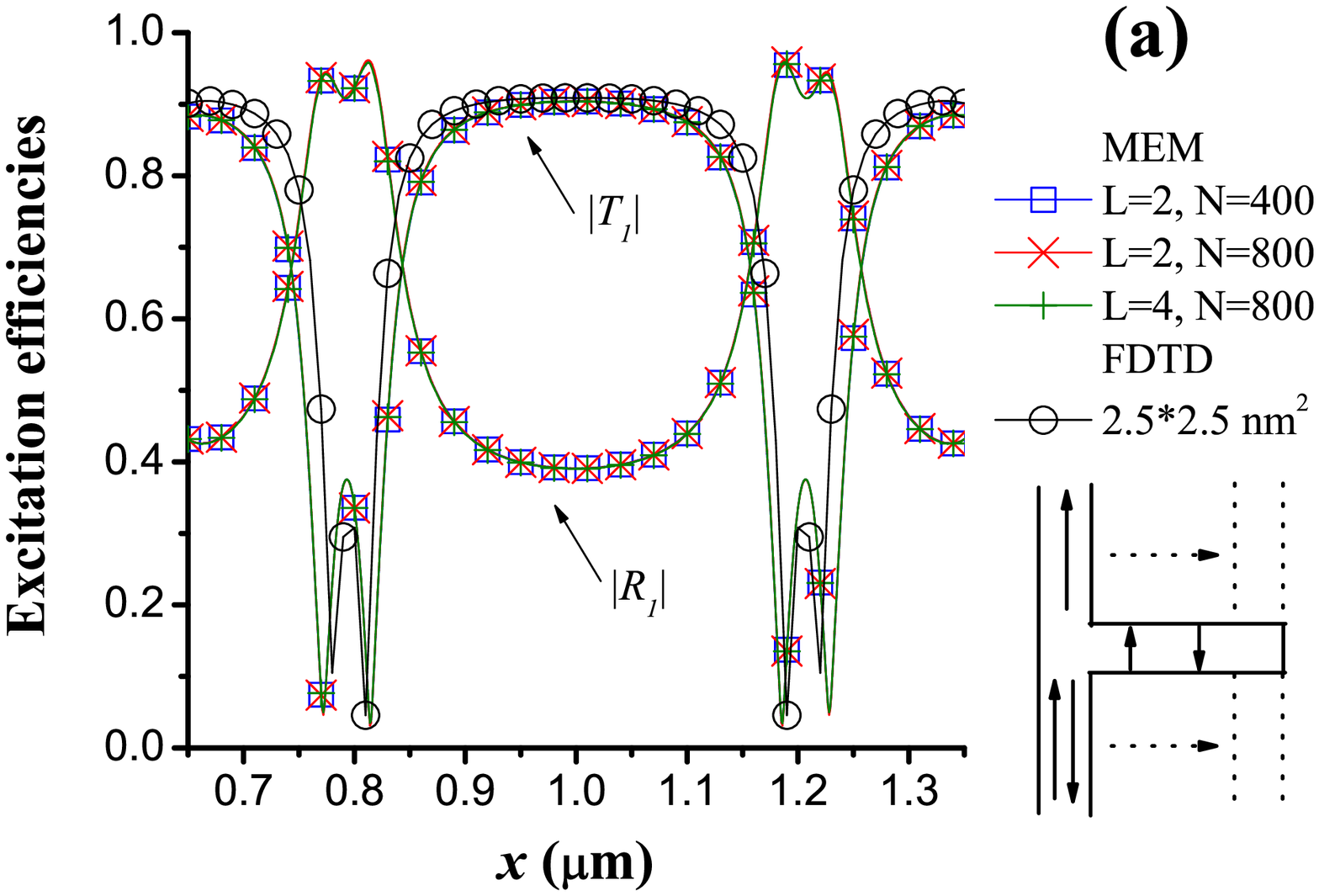}
\includegraphics  [width=6cm] {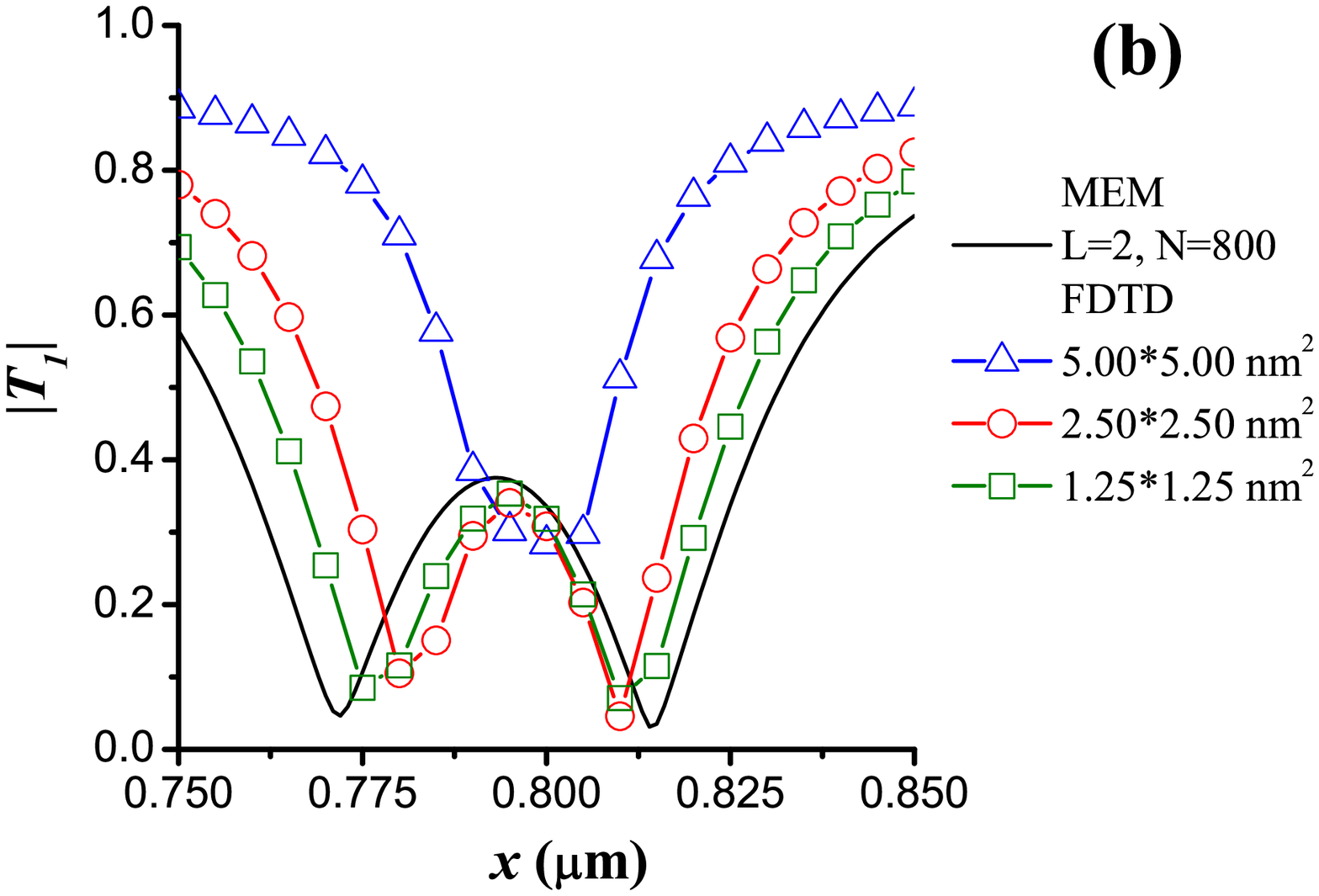}
 \caption{(color online). The SPP transmission and reflection in a Type I
structure with parameter $x_{1}^{(2)}=0.6$ $\mu$m,
$\left[w^{(1)},w^{(2)},w^{(3)}\right]=\left[0.1,0.8,0.1\right]$
$\mu$m, $\left[Q^{(0)},Q^{(1)},Q^{(2)}\right]=\left[-1,0,0.1\right]$
$\mu$m, see the inset of (a). (a) Convergence test for MEM; (b)
convergence test for FDTD relative to MEM where $x$-axis is a part
of (a) for $x_{1}^{(1)}\in\left[0.7,0.8\right]$ $\mu$m.}
\end{figure}

Figure 7(a) shows the calculated results of MEM with confined width
being $L=2,4$ $\mu$m and truncation number being $N=400,800$,
respectively. The results by squares, crosses and plus signs in Fig.
7(a) are identical, showing that the boundary effect imposed by the
perfectly conducting walls can be completely ignored in such
confined widths. As already mentioned at the beginning of Sec. 3,
the parameters $L=2$ $\mu$m and $N=800$ ensure that all the
calculated results have at least four significant digits. A FDTD
simulated transmission curve with the cell size being $2.5\times2.5$
nm$^{2}$ is also plotted in Fig. 7(a) for comparison. Obviously, the
FDTD curve is close to the MEM ones but not coincide. In order to
investigate the convergence of FDTD, three simulated transmission
curves with different cell sizes are plotted in Fig. 7(b) for the
same structure as in Fig. 7(a) but with horizontal abscissa being
$x_{1}^{(1)}\in[0.7,0.8]$ $\mu$m to highlight two absorption peaks.
A MEM curve also plotted in the figure for comparison. For large
cell size as $5\times5$ nm$^{2}$, only one vague dip, instead of
two, is observed. The dip will gradually separate into two and
approach to the MEM curve as the cell size decreases. However, even
for $1.25\times1.25$ nm$^{2}$, namely, $1/800$ of the incident
wavelength or $1/80$ of the stub width (height of Layer 2), the
deviation of the results between FDTD and MEM is still observable,
which means that FDTD has a relatively slow convergence. That is why
we do not use FDTD to verify the calculated results of MEM in this
paper.

The transmission of the structure considered in Fig. 7 was also
investigated by SMM [9]. This method actually utilizes some results
of FDTD to obtain scatting matrix elements and loses some phase
information by quasi-statistic approximation, so that its final
results could not be better than that of FDTD. This can be
recognized by the comparison between the results of FDTD and SMM
given in Refs. [8] and [9]. Therefore, it is not necessary to
discuss the preciseness of SMM here because it depends on the
simulation results of FDTD which has already shown in Fig. 7(b).
Besides, since the SMM and ICIM have the similar transmission model,
their calculation errors ought to have the same order of magnitude.
The calculation error of ICIM is investigated in the following.

The SPP transmission calculated by ICIM and MEM in a Type II
structure are plotted in Figs. 8 and 9. This kind of structure was
also study in Refs. [13] and [14].

\begin{figure}[htbp]
\centering\includegraphics  [width=6.5cm] {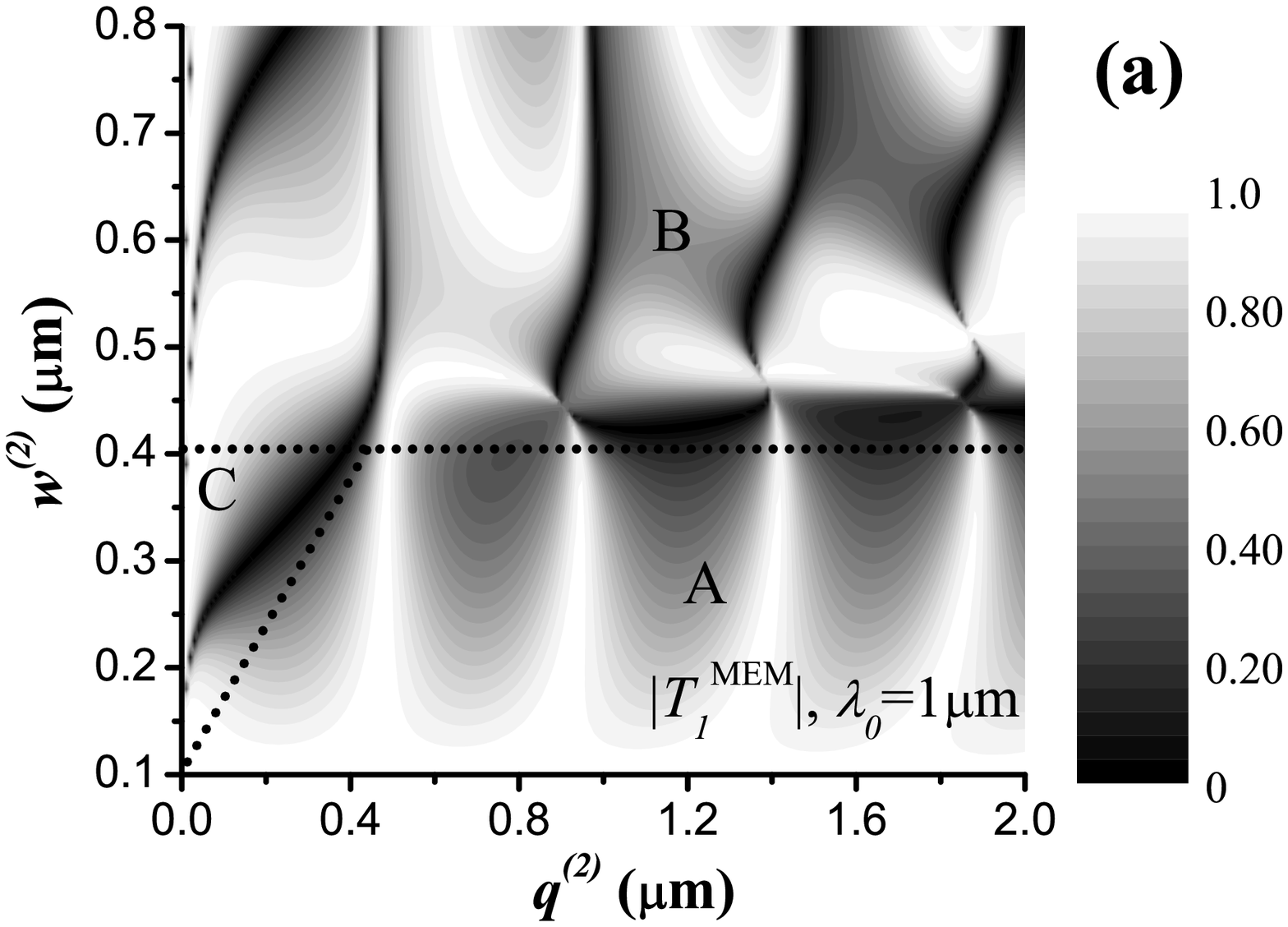}
\includegraphics  [width=6.5cm] {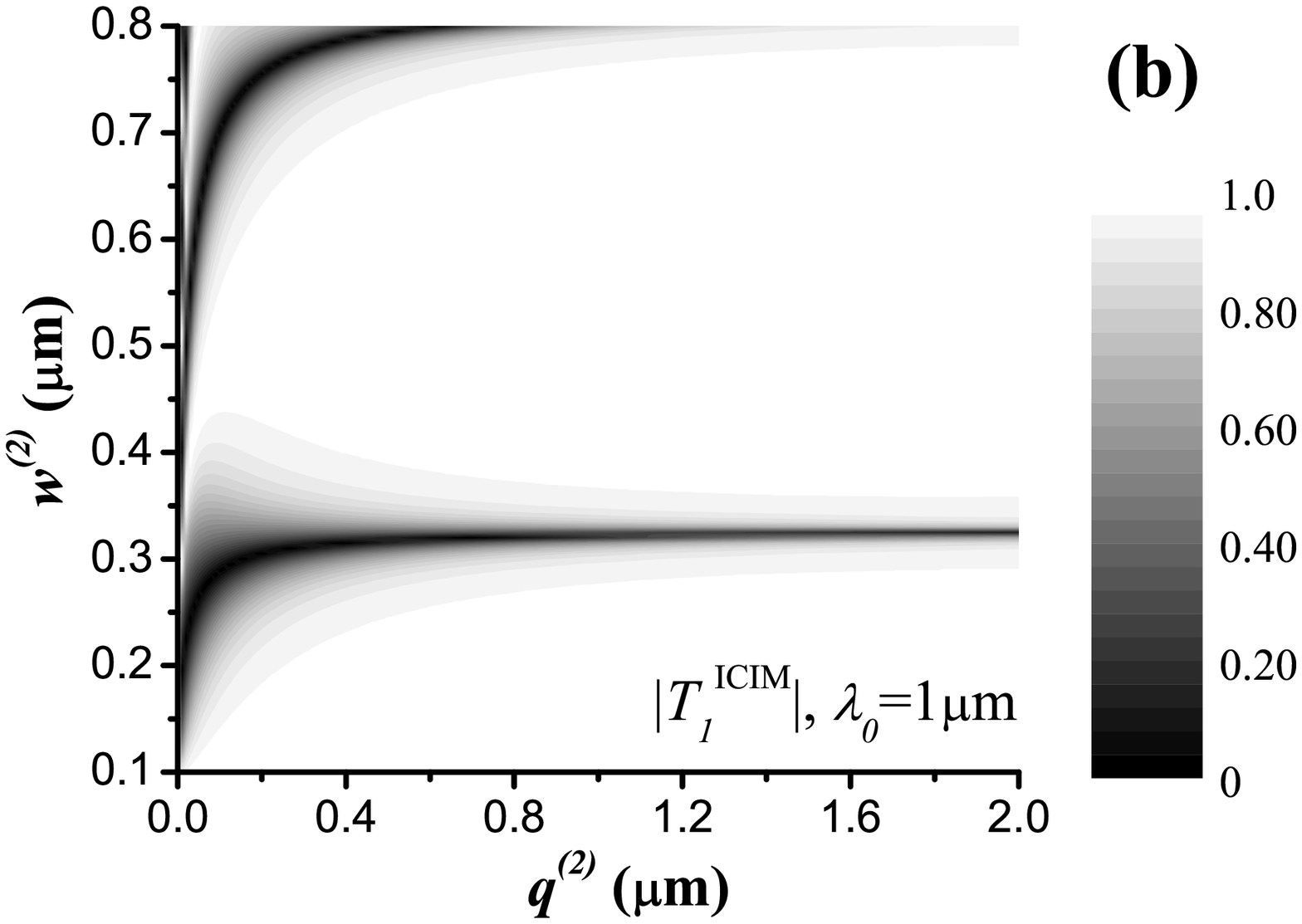}
\includegraphics  [width=6.5cm] {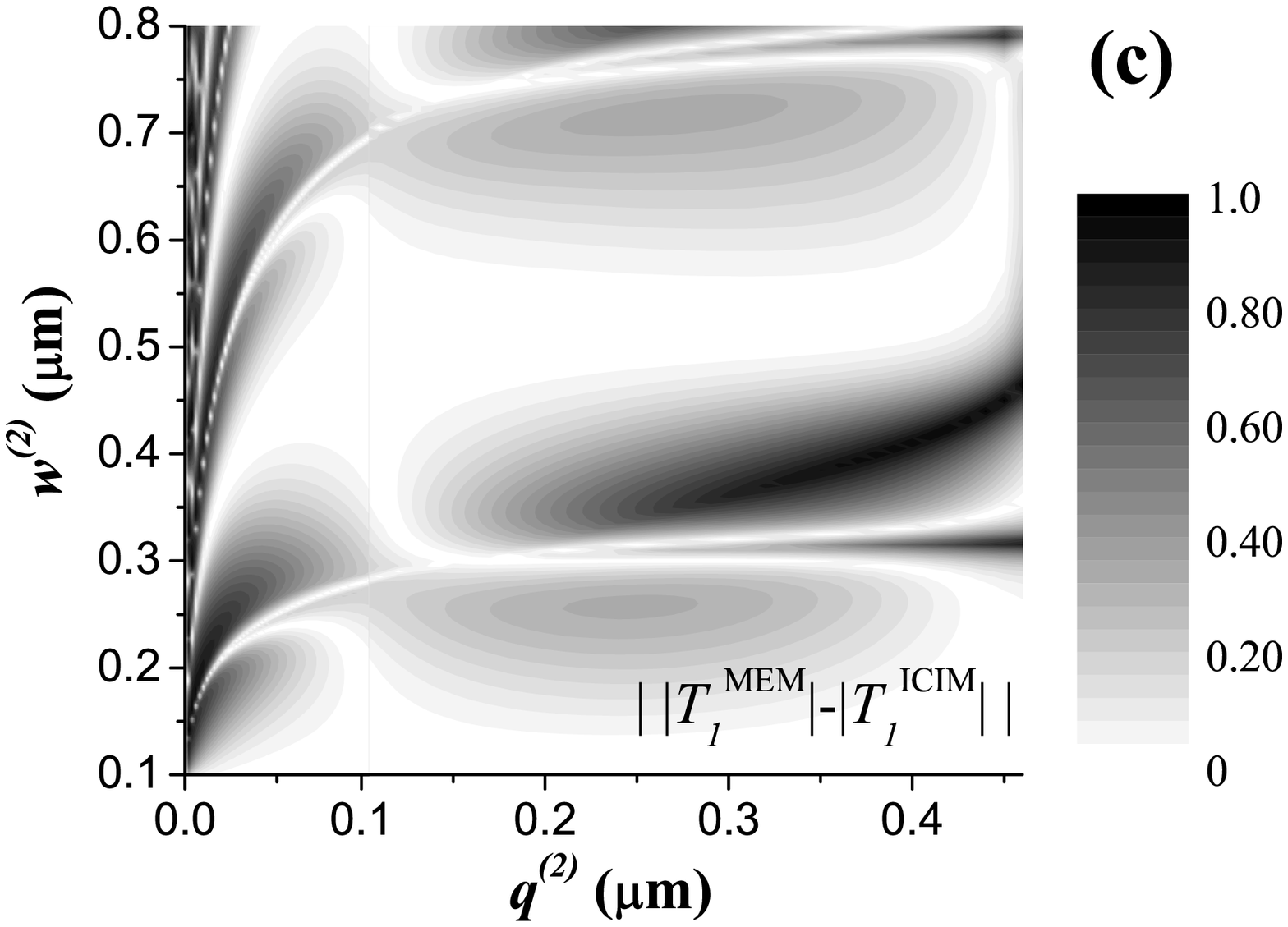}
 \caption{Comparison of the SPP transmission by MEM and ICIM of a Type II
structure under the variation of $w^{(2)}$ and $q^{(2)}$. The left
sides of the slits in all layers are aligned to $x_{1}^{(l)}=0.6$
$\mu$m.
$\left[w^{(1)},w^{(2)},w^{(3)}\right]=\left[0.1,w^{(2)},0.1\right]$
$\mu$m,
$\left[Q^{(0)},Q^{(1)},Q^{(2)}\right]=\left[-1,0,q^{(2)}\right]$
$\mu$m. (a) $|T_{1}^{MEM}|$ by MEM; (b) $|T_{1}^{ICIM}|$ by ICIM;
(c) absolute value of the difference between MEM and ICIM,
$\left||T_{1}^{MEM}|-|T_{1}^{ICIM}|\right|$ for $q^{(2)}<0.46$
$\mu$m.}
\end{figure}

In Fig. 8 are given the SPP transmission as a function of the height
and width of Layer 2 under a fixed incident wavelength,
$\lambda_{0}=1$ $\mu$m. In Fig. 8(a), the calculated results of MEM
can be approximately divided into three areas A, B, and C by dotted
lines. In Area A, a regular oscillating pattern is observed because
when $w^{(2)}<0.4$ $\mu$m only one SPP mode propagates in Layer 2
that forms the FP-like oscillation [5,6]. While in Areas B and C
where the higher modes also contribute to transmission, the
transmission pattern becomes complicated. Intuitively, there are two
ways for higher modes to transport energy. One is in a way of a
propagation mode, which is appropriate for $w^{(2)}>0.46$ $\mu$m
because the 2nd mode in Layer 2 becomes propagating. The other is in
a way of an evanescent mode, which is appropriate for
$0.4<w^{(2)}<0.46$ $\mu$m and Area C because the 2nd mode in Layer 2
does not attenuate to a negligible value in such a modulated slit
and brings energy through Layer 2.

The results from ICIM shown in Fig. 8(b) can be discussed according
to if $q^{(2)}$ is larger or less than $0.46$ $\mu$m. When
$q^{(2)}>0.46$ $\mu$m, more than one mode are allowed to propagate
in the stub, while the ICIM assumes only the SPP mode, which raises
the great difference relative to the MEM results, leading to totally
different patterns between Figs. 8(a) and (b). When $q^{(2)}<0.46$
$\mu$m, the difference $\left||T_{1}^{MEM}|-|T_{1}^{ICIM}|\right|$
is plotted in Fig. 8(c). In this region, the difference is mainly
caused by the neglect of the phase shifts in the central scattering
region, which indicates that these phase shifts have to be taken
into account in calculation. In Fig. 8(c), it is seen that when the
length of Layer 2 $q^{(2)}<0.03$ $\mu$m, less than $1/30$ of the
incident wavelength, the difference becomes larger due to the energy
transported by evanescent modes. This demonstrates that a very
narrow region cannot guarantee precise results. Thus, the
application of the quasi-statistic approximation in a modulated
metal slit requires an optimum geometry in order to provide
relatively accurate results: the incident wavelength is nearly 10
times larger than stub width.

\begin{figure}[htbp]
\centering\includegraphics  [width=6.5cm] {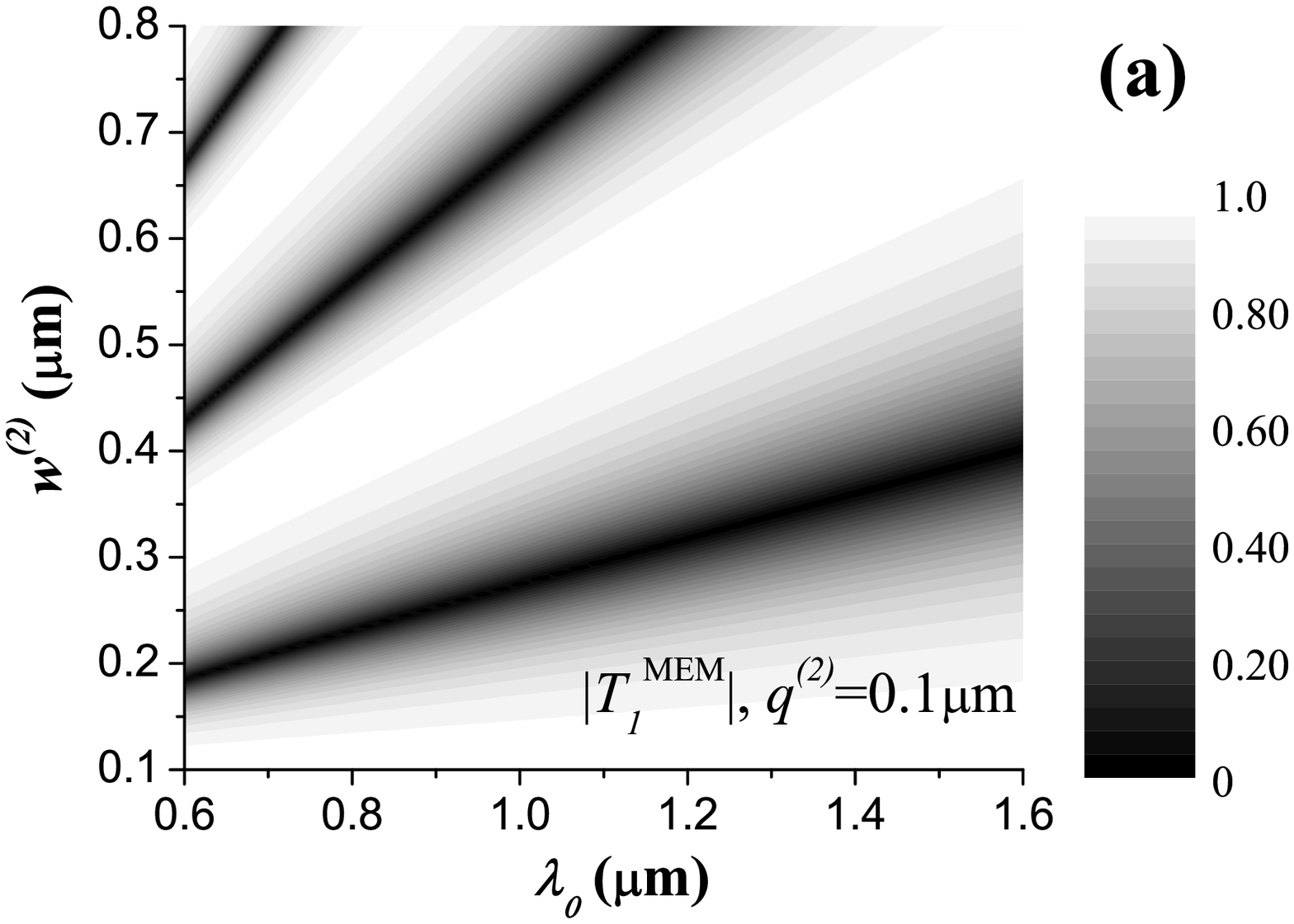}
\includegraphics  [width=6.5cm] {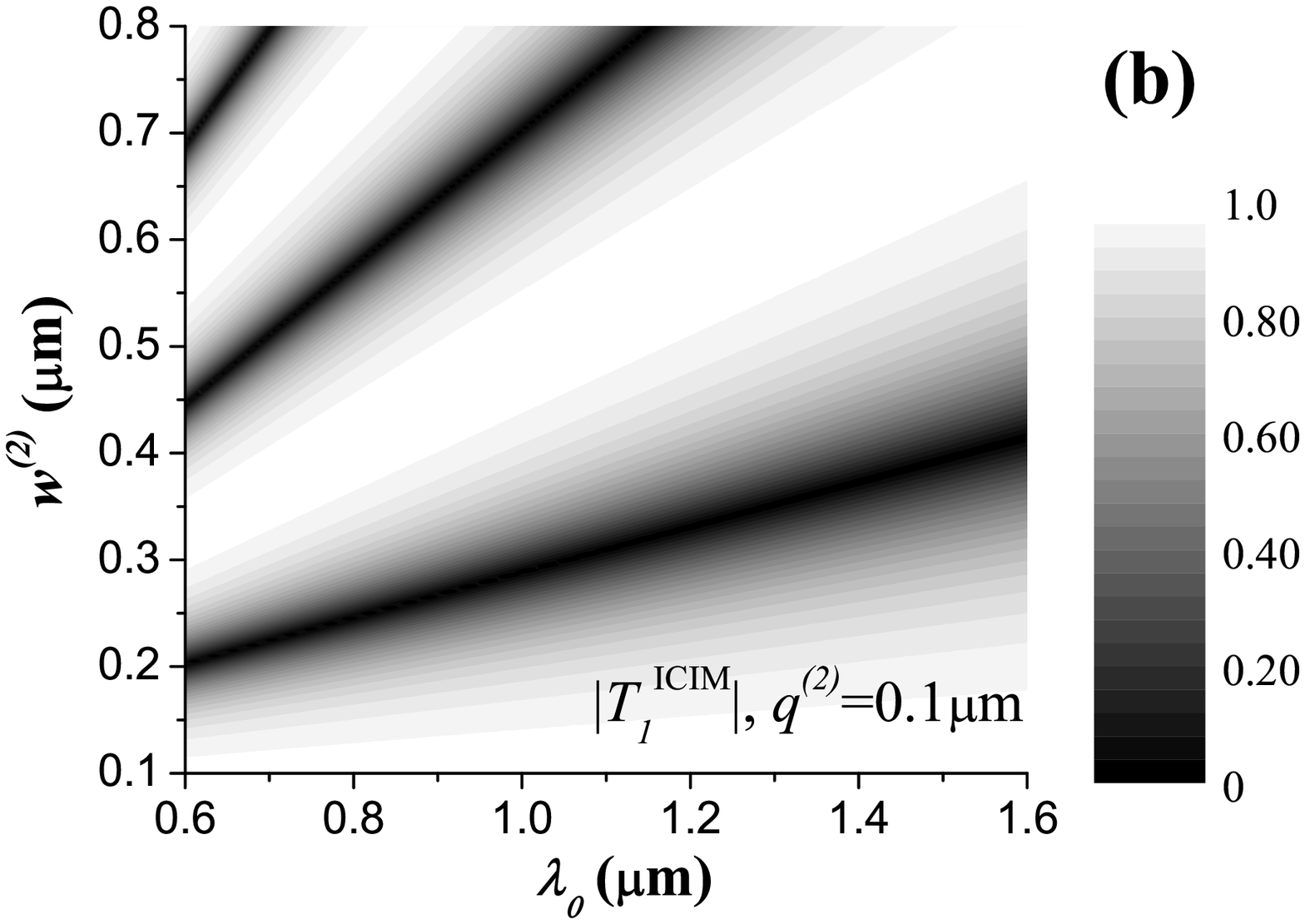}
\includegraphics  [width=6.5cm] {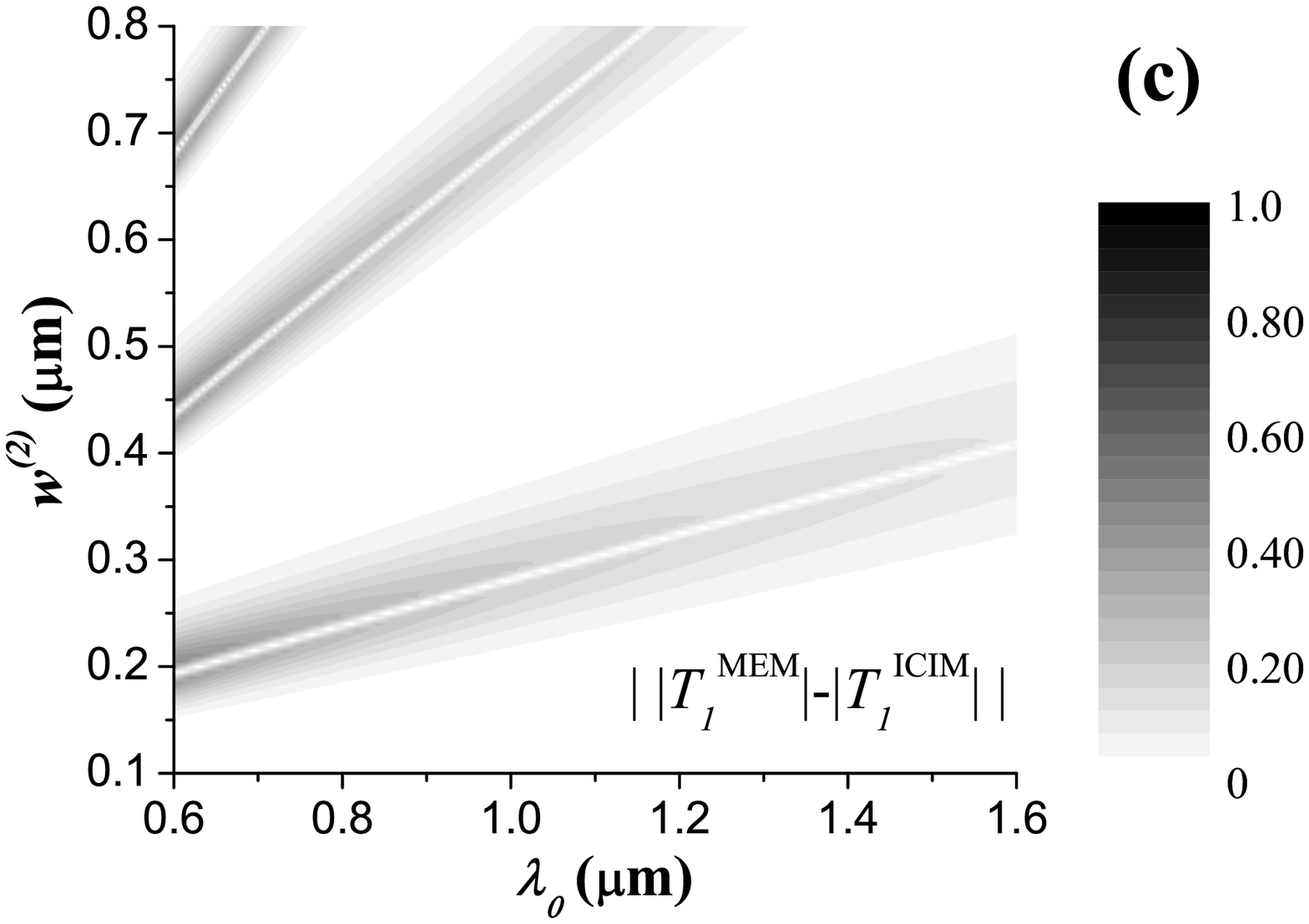}
 \caption{Comparison of the SPP transmission by MEM and ICIM of a Type II
structure under the variation of $w^{(2)}$ and wavelength
$\lambda_{0}$. The left sides of the slits in all layers are aligned
to $x_{1}^{(l)}=0.6$ $\mu$m.
$\left[w^{(1)},w^{(2)},w^{(3)}\right]=\left[0.1,w^{(2)},0.1\right]$
$\mu$m, $\left[Q^{(0)},Q^{(1)},Q^{(2)}\right]=\left[-1,0,0.1\right]$
$\mu$m. (a) $|T_{1}^{MEM}|$ by MEM; (b) $|T_{1}^{ICIM}|$ by ICIM;
(c) absolute value of the difference between MEM and ICIM,
$\left||T_{1}^{MEM}|-|T_{1}^{ICIM}|\right|$.}
\end{figure}

In order to test the optimum geometry, the slit width of Layer 2 and
the incident wavelength $\lambda_{0}$ are considered as variables
and the length of Layer 2 is set as  $q^{(2)}=0.1$ $\mu$m, the same
as the slit widths in Layers 1 and 3. The MEM results in Fig. 9(a)
exhibit a few straight black bands indicating the
inverse-proportional relation between frequency and stub width which
was mentioned in Refs. [8] and [9]. For ICIM, similar results are
obtained in Fig. 9(b). The difference of these two figures is
plotted in Fig. 9(c). Clearly, the calculation error of ICIM in this
case is lower than that in Fig. 8(c). Although small deviation takes
place near the absorption peaks, ICIM successfully predicts the
positions of the peaks. All these numerical results confirm that the
ICIM can provide relatively accurate results when the incident
wavelength is around 10 times of the stub width.

Anyway, each method has its own advantages in certain aspect. For
example, FDTD can provide visualized transmission process in time
domain and ICIM is indubitably the fastest method in calculation.
Here we just emphasize that these methods should be used with
caution in considering the convergence and preciseness.

\section{Conclusion}

In this paper, the MEM developed in Ref. [6] is improved to be a
more practical and efficient one for handling the
scattering/transmission. Using this method, the scattering in a
juncture structure and the transmission inside a step-modulated slit
are investigated.

Firstly, the scattering in a juncture structure is studied for two
types of structural changes. For the Type I change where the widths
of the two slits are fixed and the position of the narrower one can
be anywhere within the wider one, the excitation efficiencies of the
modes in the wider slit resemble their eigenfunctions respectively,
while in the narrower slit the excitation efficiency of the SPP mode
is dominant and much larger than that of the second mode. For the
Type II change where the left walls of the slits are aligned and one
slit becomes wider gradually, a wood-anomaly-like drastic change of
excitation efficiencies is observed when the propagation property of
one mode begins to transform.

Then, the transmission inside a step-modulated slit is studied.
Besides the MEM calculation, we present explicitly a multi-mode
multi-reflection model to reveal the transmission process. The
multi-mode excitation and the superposition procedure of the
scatterings from all possible modes are the key parts of the model,
which represent the interference and energy transfer happened at
layer boundaries. However, there exists a critical height of the
modulated layer for applying the model due to the coupling between
propagation modes and evanescent modes. Above the critical height,
the model can provide the same result as MEM.

In addition, some commonly used methods are compared with MEM.
Useful conclusions about these methods are obtained: for a
subwavelength metal slit, MEM is a versatile and fast method that
can provide accurate results; FDTD has a relatively slow convergence
and need very small Yee cell to ensure the accuracy of the
simulation; ICIM incorporating the one mode and quasi-statistic
approximations provides relatively accurate results when the
incident wavelength is around 10 times larger than the stub width.

\section*{Acknowledgments}

This work is supported by the 973 Program of China (Grant
No.2011CB301801) and the National Natural Science Foundation of
China (Grant No. 10874124), and (Grant No. 11074145).

\end{document}